%% file: main.tex
\documentclass[journal]{IEEEtran}

\usepackage{lipsum}
\usepackage[T1]{fontenc}

\usepackage{latexsym}
\usepackage{graphicx}
\usepackage{tabularx}
\usepackage{multirow}
\usepackage{booktabs}
\graphicspath{{img/}}
\usepackage{amsfonts,amssymb,amsmath,systeme}
\usepackage[hidelinks]{hyperref}
\usepackage[capitalise]{cleveref}
\usepackage{tikz}
\usetikzlibrary{shapes, arrows, positioning, fit, calc, backgrounds}
\usepackage[american]{circuitikz}

\tikzset{
  pics/busbar/.style n args={3}{
    code={
      \def\buslen{#1}
      \def\abovetext{#2}
      \def\belowtext{#3}
      \coordinate (-center) at (0,0);
      \coordinate (-north)  at (0,\buslen);
      \coordinate (-south)  at (0,-\buslen);
      \draw[ultra thick] (-north) -- (-south);
      \node[anchor=base] at ($(-north)+(0,0.15)$) {\abovetext};
      \node[below]        at ($(-south)-(0,0.15)$) {\belowtext};
    }
  }
}

\newcommand{\bushere}[3]{pic {busbar={#1}{#2}{#3}}}
\newcommand{\busherehor}[3]{% length, text above, text below
    coordinate(tmp)
    ++(#1,0) node[right]{#2} edge[ultra thick] ++({-2*#1},0)
    ++({-2*#1},0) node[above]{#3}
    (tmp)
}

\usepackage{soul}

\usepackage{pgfplots}
\pgfplotsset{compat=newest}
\pgfplotsset{plot coordinates/math parser=false}
\newlength\figureheight
\newlength\figurewidth
\usepackage{eurosym}
\usepackage{xcolor}
\usepackage{float}
\usepackage{dblfloatfix}
\usepackage{caption}
\usepackage{subcaption}
\usepackage{enumitem}
\usepackage{multicol}
\usepackage{cite}
\usepackage{todonotes}
\definecolor{kit-green}{RGB}{0, 150, 130}
\definecolor{kit-blue}{rgb}{0.27451,0.39216,0.66667}%
\definecolor{mycolor2}{rgb}{0.00000,0.58824,0.50980}%
\definecolor{kit-yellow}{rgb}{0.95294,0.61176,0.07059}%
\definecolor{kit-red}{rgb}{0.90588,0.29804,0.23529}%

\newcommand\footnoteref[1]{\protected@xdef\@thefnmark{\ref{#1}}\@footnotemark}

\usepackage[symbols, nonumberlist, nogroupskip, nopostdot]{glossaries}

\glsdisablehyper

\makeglossaries

\loadglsentries{glossarentries.tex}
\loadglsentries[type=symbols]{glossarentries_symbols.tex}

\definecolor{lime}{HTML}{A6CE39}
\DeclareRobustCommand{\orcidicon}{
	\begin{tikzpicture}
	\draw[lime, fill=lime] (0,0) 
	circle [radius=0.16] 
	node[white] {{\fontfamily{qag}\selectfont \tiny ID}};
	\draw[white, fill=white] (-0.066,0.09) 
	circle [radius=0.003];
	\end{tikzpicture}
	\hspace{-2mm}
}

\foreach \x in {A, ..., Z}{\expandafter\xdef\csname orcid\x\endcsname{\noexpand\href{https://orcid.org/\csname orcidauthor\x\endcsname}
			{\noexpand\orcidicon}}
}

\def\BibTeX{{\rm B\kern-.05em{\sc i\kern-.025em b}\kern-.08em T\kern-.1667em\lower.7ex\hbox{E}\kern-.125emX}}

\begin{document}

%\pgfplotsset{plot coordinates/math parser=false}
\title{Asynchronous Grid Connections Providing Fast-Frequency Response: System Integration Study}

%%%% Authors
\author{Felix Wald$^{\orcidA}${}  \IEEEmembership{Student Member, IEEE}, Amir Sajadi$^{\orcidB}${}  \IEEEmembership{Senior Member, IEEE},\\ Barry Mather$^{\orcidC}${}  \IEEEmembership{Senior Member, IEEE}, and Giovanni De Carne$^{\orcidD}${}  \IEEEmembership{Senior Member, IEEE}
%\thanks{This work was supported by the Helmholtz Association under the program "Energy System Design" and the Helmholtz Young Investigator Group "Hybrid Networks" (VH-NG-1613).}
\thanks{Felix Wald and Giovanni De Carne are with the Institute for Technical Physics at Karlsruhe Institute of Technology, Karlsruhe, Germany (e-mail: felix.wald@kit.edu).}

\thanks{Amir Sajadi is with the University of Colorado Boulder, Boulder, CO 80309, USA, and the Power Systems Engineering Center at the National Laboratory of the Rockies, Golden, CO 80401, USA.}
\thanks{Barry Mather is with the Power Systems Engineering Center at the National Laboratory of the Rockies, Golden, CO 80401, USA.}
%\thanks{Corresponding Author: felix.wald@kit.edu }
}

\markboth{ASG Integration Study for IEEE Transactions on Power Delivery, January 2026}%
{Shell \MakeLowercase{\textit{et al.}}: Bare Demo of IEEEtran.cls for IEEE Journals}

\maketitle

\begin{abstract}
This paper presents an integration study for the recent power electronic-based fast-frequency response technology, "asynchronous grid connection" which operates as an aggregator for behind-the-meter resources and distributed generators. Both technical feasibility and techno-economic viability studies are presented. The fast-frequency response characteristics, validated against Power Hardware-in-the-Loop experiments, are integrated into an IEEE 9-bus system in DigSilent PowerFactory for system-level dynamic analysis. It demonstrates that droop-based control enhancements to local distributed generators allow their aggregation to provide grid-supporting functionalities and participate in the ancillary service markets. To this end, a long-term simulation embedding the system within the ancillary service market framework of PJM has been performed. The fast-frequency response regulation is subsequently used to calculate the potential revenue and project the results on a 15-year investment horizon. Finally, the techno-economic analysis provides recommendations for enhancements to access the full potential of distributed generators on a technical and regulatory level.
\end{abstract}


%%%%% Glossary 
%this algins the full name depending on the lablewidth, choose accodring to the longest abbreviation
%\setlist[description]{leftmargin=!, labelwidth=5em, font=\normalfont\space} %set the label width, this aligns the fullform with that width, change the abrreviation font style to normal instead of bold
%\printglossary[title=Acronyms] % defines the title, uses the standard style, here the option "style=super" could be used to use a better looking style, but it does not work with the double column IEEE clasee
%% Separate list for the math symbols and variabels
%\printglossary[type=symbols]
%\setlist[description]{style=standard} % resets the style for the rest of the document

\glsresetall

%%%%% Content Files
\input{sections/01_introduction}
\input{sections/02_modeling}

\input{sections/03_transient_analysis}
\input{sections/04_economics}

\input{sections/05_discussion}
\input{sections/06_conclusion}
\input{sections/yy_appendix}

\section*{Acknowledgment}
We wish to thank Enrique Bacalao with the Bacalao Consulting Services, LLC, for his insightful discussions and comments.

This work was supported by the Helmholtz Association under the program "Energy System Design" and the Helmholtz Young Investigator Group "Hybrid Networks" (VH-NG-1613).

This work was authored in part by the National Laboratory of the Rockies for the U.S. Department of Energy (DOE), operated under Contract No. DE-AC36-08GO28308. Funding provided by U.S. Department of Energy Office of Energy Efficiency and Renewable Advanced Materials and Manufacturing Technologies Office. The views expressed in the article do not necessarily represent the views of the DOE or the U.S. Government. The U.S. Government retains and the publisher, by accepting the article for publication, acknowledges that the U.S. Government retains a nonexclusive, paid-up, irrevocable, worldwide license to publish or reproduce the published form of this work, or allow others to do so, for U.S. Government purposes.
\ifCLASSOPTIONcaptionsoff
  \newpage
\fi

\bibliographystyle{IEEEtran}
\bibliography{bibtex/bib/Full_Library_bibtex, bibtex/bib/manual_bib}

%
\input{sections/zz_bios}

\end{document}

%% file: glossarentries.tex
%A
\newacronym{AFE}{AFE}{active front end}
\newacronym{AGC}{AGC}{Automatic Generation Control}
\newacronym{ACE}{ACE}{Area Control Error}
\newacronym{ASG}{ASG}{Asynchronous Grid}

%B
\newacronym{B2B}{B2B}{Back-to-Back Converter}
\newacronym{BESS}{BESS}{Battery Energy Storage System}

%C
\newacronym{CHIL}{CHIL}{Controller Hardware-in-the-Loop}
\newacronym{CIG}{CIG}{Converter-Interfaced Generation}

%D
\newacronym{DER}{DER}{Distributed Energy Resource}
\newacronym{DG}{DG}{Distributed Generation}
\newacronym{DSR}{DSR}{Demand Side Response}

%E
\newacronym{ESS}{ESS}{Energy Storage System}
\newacronym{EMI}{EMI}{electromagnetic interference}
\newacronym{ENTSO-E}{ENTSO-E}{European Network of Transmission System Operators for Electricity}

%F
\newacronym{FFR}{FFR}{Fast-Frequency Response}
\newacronym{FERC}{FERC}{Federal Energy Regulatory Commission}
\newacronym{FCR}{FCR}{Frequency Containment Reserve}

%G
\newacronym{GFL}{GFL}{Grid-Following}
\newacronym{GFM}{GFM}{Grid-Forming}

%H
\newacronym{HIL}{HIL}{Hardware-in-the-Loop}

%I
\newacronym{ITEP}{ITEP}{Institute for Technical Physics}
\newacronym{ISO}{ISO}{International Organization for Standardization}
\newacronym{IRENA}{IRENA}{International Renewable Energy Agency}
\newacronym{IRR}{IRR}{Internal Rate of Return}

%J

%K
\newacronym{KVL}{KVL}{Kirchhoffs Voltage Law}

%L
\newacronym{LFT}{LFT}{Line-frequency Transformer}
\newacronym{LVRT}{LVRT}{Low-Voltage Ride-through}

%M
\newacronym{MMC}{MMC}{Modular Multi-level Converter}
\newacronym{MOSFET}{MOSFET}{metal-oxide-semiconductor field-effect transistor}

%N
\newacronym{NPV}{NPV}{Net Present Value}
\newacronym{NWA}{NWA}{Non-wire Alternativ}
%O

%P
\newacronym{PCC}{PCC}{Point of Common Coupling}
\newacronym{PHIL}{PHIL}{Power Hardware-in-the-Loop}
\newacronym{PI}{PI}{proportional integral}
\newacronym{PR}{PR}{proportional resonant}
\newacronym{PV}{PV}{photovoltaic}
\newacronym{pu}{\text{pu}}{\text{per unit}}
\newacronym{PLL}{PLL}{phase-looked loop}
\newacronym{PWM}{PWM}{Pulse Width Modulation}

%Q

%R
\newacronym{RES}{RES}{Renewable Energy Source}
\newacronym{RoCoF}{RoCoF}{Rate of Change of Frequency}
\newacronym{RMS}{RMS}{Root-Means-Square}
\newacronym{RTO}{RTO}{Regional Transmission Organization}

%S
\newacronym{SoC}{SoC}{State-of-Charge}
\newacronym{SCL}{SCL}{Short-Circuit Level}
\newacronym{SG}{SG}{Synchronous Generator}
\newacronym{SiC-MOSFET}{SiC-MOSFET}{silicon carbide metal-oxide-semiconductor field-effect transistor}
\newacronym{SM}{SM}{Synchronous Machine}
\newacronym{SST}{SST}{Solid-State Transformer}

%T
\newacronym{TSO}{TSO}{Transmission System Operator}

%V
\newacronym{V2G}{V2G}{Vehicle-to-Grid}
\newacronym{VPP}{VPP}{Virtual Power Plant}

\newacronym{VRE}{VRE}{Variable Renewable Energy}
\newacronym{VSM}{VSM}{Virtual Synchronous Machine}
\newacronym{VSC}{VSC}{Voltage-source Converter}

%% file: glossarentries_symbols.tex
\newglossaryentry{Mratio}{
  type={symbols},
  name={\ensuremath{\beta_{\text{t}}^{\text{M}}}},
  description={Mileage ratio between RegD and RegA},
  sort={Mratio}
}
\newglossaryentry{Cashflow}{
  type={symbols},
  name={\text{Cf}},
  description={Cash flow},
  sort={Cashflow}
}
\newglossaryentry{Creg}{
  type={symbols},
  name={\ensuremath{\text{C}_{\text{reg}}}},
  description={Regulation Credit},
  sort={Creg}
}
\newglossaryentry{Dp}{
  type={symbols},
  name={\ensuremath{\text{D}_{\text{p}}}},
  description={VSM damping coefficient},
  sort={Dp}
}
\newglossaryentry{Gasg}{
  type={symbols},
  name={\ensuremath{\text{G}_{\text{ASG}}}},
  description={Asynchronous grid controller transfer function},
  sort={Gasg}
}
\newglossaryentry{Ggov}{
  type={symbols},
  name={\ensuremath{\text{G}_{\text{gov}}}},
  description={Governor controller transfer function},
  sort={Ggov}
}
\newglossaryentry{Glv}{
  type={symbols},
  name={\ensuremath{\text{G}_{\text{LV}}}},
  description={LV distribution grid transfer function},
  sort={Glv}
}
\newglossaryentry{Gprop}{
  type={symbols},
  name={\ensuremath{\text{G}_{\text{prop}}}},
  description={Frequency propagation gain},
  sort={Gprop}
}
\newglossaryentry{Gvsm}{
  type={symbols},
  name={\ensuremath{\text{G}_{\text{vsm}}}},
  description={VSM transfer function},
  sort={Gvsm}
}
\newglossaryentry{interest}{
  type={symbols},
  name={\text{r}},
  description={Interest rate},
  sort={interest}
}
\newglossaryentry{InvCost}{
  type={symbols},
  name={\text{I}},
  description={Invesment Cost},
  sort={InvCost}
}
\newglossaryentry{Kpf_scaling}{
  type={symbols},
  name={\alpha},
  description={Scaling Kpf to LV composition},
  sort={alpha}
}
\newglossaryentry{Kigov}{
  type={symbols},
  name={\ensuremath{\text{K}_{\text{igov}}}},
  description={Integral gain - Governor controller},
  sort={Kigov}
}
\newglossaryentry{Kpf}{
  type={symbols},
  name={\ensuremath{\text{K}_{\text{pf}}}},
  description={Active power to frequency sensitivity},
  sort={Kpf}
}
\newglossaryentry{Kpgov}{
  type={symbols},
  name={\ensuremath{\text{K}_{\text{pgov}}}},
  description={Proportional gain - Governor Controller},
  sort={Kpgov}
}
\newglossaryentry{RCCP}{
  type={symbols},
  name={\ensuremath{\lambda_{\text{C}}}},
  description={Regulation Capability Clearing Price},
  sort={RCCP}
}
\newglossaryentry{RPCP}{
  type={symbols},
  name={\ensuremath{\lambda_{\text{P}}}},
  description={Regulation Performance Clearing Price},
  sort={RPCP}
}
\newglossaryentry{MRegA}{
  type={symbols},
  name={\ensuremath{\text{M}_{\text{RegA}}}},
  description={Mileage of RegA signal},
  sort={MRegA}
}
\newglossaryentry{MRegD}{
  type={symbols},
  name={\ensuremath{\text{M}_{\text{RegD}}}},
  description={Mileage of RegD signal},
  sort={MRegD}
}
\newglossaryentry{OM}{
  type={symbols},
  name={\ensuremath{\text{c}_{\text{OM}}}},
  description={Operation and Maintenance Cost},
  sort={OM}
}
\newglossaryentry{para_vec}{
  type={symbols},
  name={\ensuremath{\mathbf{p}}},
  description={AG Support model parameter vector},
  sort={p}
}
\newglossaryentry{P0}{
  type={symbols},
  name={\ensuremath{P_{0}}},
  description={Rated active power},
  sort={P0}
}
\newglossaryentry{Pasg}{
  type={symbols},
  name={\ensuremath{P_{\text{ASG}}}},
  description={Support active power of the asynchronous grid},
  sort={Pasg}
}
\newglossaryentry{Pgov}{
  type={symbols},
  name={\ensuremath{P_{\text{gov}}}},
  description={Governor active power set point},
  sort={Pgov}
}
\newglossaryentry{Pff}{
  type={symbols},
  name={\ensuremath{P_{\text{ff}}}},
  description={Propagated feed-forward active power set point},
  sort={Pff}
}
\newglossaryentry{RegD}{
  type={symbols},
  name={\text{RegD}},
  description={Regulation in one time step},
  sort={RegD}
}
\newglossaryentry{Rt}{
  type={symbols},
  name={\ensuremath{\text{R}_{\text{t}}}},
  description={Net cash flow in year t},
  sort={Rt}
}
\newglossaryentry{Ta}{
  type={symbols},
  name={\ensuremath{\text{T}_{\text{a}}}},
  description={Virtual mechanical torque of the VSM},
  sort={Ta}
}

%% file: sections/01_introduction.tex
\section{Introduction}

\IEEEPARstart{T}{he} transition towards \glspl{RES} is crucial, as the \gls{IRENA} advocates for a 91\% share in electricity generation by 2050~\cite{irena2023WorldEnergyTransitions2023,irena2020GlobalRenewablesOutlook2020}. This shift presents new challenges, particularly in maintaining frequency stability, historically managed by synchronous generators~\cite{milanoFoundationsChallengesLowinertia2018,sajadi2022synchronization}. The replacement of fossil fuel-based plants is not only necessary to meet the emission targets but also economically viable, given the increasing cost competitiveness of \glspl{RES}~\cite{Lazard2024}.
Of the technological barriers to realize this transition, this paper addresses the challenge of providing reliable and efficient \gls{FFR} services~\cite{sajadi2022electric}.
Although regional system operators may recognize this service under different market terms, rules, and mechanisms, for ease of reference, in the remainder of this text it is referred to as \gls{FFR}. This applies when the service is delivered within 10 seconds of an event or contingency in order to participate in the respective service framework and subsequently become entitled to compensation. 

In the past two decades, various technologies have emerged, based on controllable power electronics, as tangible solutions for providing such capabilities; specifically, converter-interfaced \glspl{RES} have demonstrated significant advantages~\cite{sajadi2022synchronization}. Implementing \gls{FFR} through \glspl{RES} requires minimal control modifications, such as embedded droop characteristics during frequency events, though this may impact economic returns through power output reductions. Moreover, support during under-frequency events necessitates continuous suboptimal operation of \glspl{RES}, significantly affecting economic viability~\cite{alsharifFastFrequencyResponse2023}.

Among the feasible solutions for \gls{FFR} services, the use of low density, behind meter resources to support bulk power systems is gaining popularity and the utilities continue to implement and launch \gls{DSR} programs~\cite{jainGridSupportiveLoadsNew2022}. Promising control strategies in the field utilize the control margins of refrigerator fleets, selectively disconnect household appliances during emergencies, or exploit load sensitivities for adaptive  power set point changes~\cite{misaghianFastFrequencyResponse2022,xuDemandFrequencyControlled2011,taoPotentialFrequencybasedPower2022}. This requires a well-developed infrastructure for direct integration into a hierarchical control structure, or a sufficiently accurate network model for indirect control purposes~\cite{prakashFastFrequencyControl2023,geis-schroerPowerFrequencyDependencyResidential2024, taoExperimentalInvestigationPowerVoltage2024, courcellePerturbationBasedLoadSensitivity2024}. Moreover, while the use of Vehicle-to-Grid offers additional storage capacity, it is at risk of low acceptance by the end user due to accelerated battery degradation~\cite{alsharifFastFrequencyResponse2023}. A significant impediment to the extensive adoption of \gls{DSR} is that a large number of devices must cooperate to provide a sufficient magnitude of power and allow participation in ancillary markets. 

\begin{figure}[t]
\centering
\includegraphics[width=.9\columnwidth]{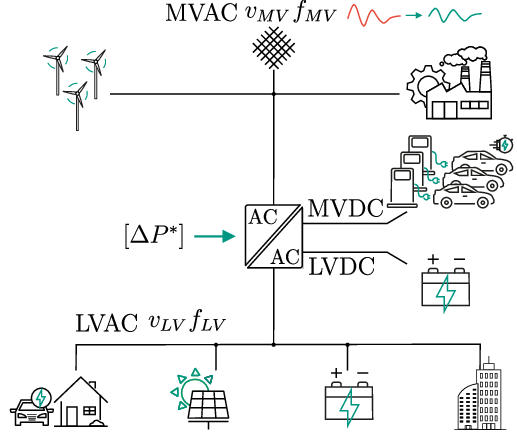}
\caption{Conceptual visualization for the asynchronous grid introduction. \label{fig:asynch-concept}}
\end{figure}

While \glspl{VPP} offer promising opportunities for distributed resource cooperation and subsequent participation in energy markets~\cite{esfahaniStochasticrobustAggregationStrategy2024, liRobustOptimizationApproach2022,yinEnergyManagementAggregate2020}, several challenges remain. For extended time horizons, decision tree-aided load regulation successfully mitigates potential over-frequency events~\cite{moutisDecisionTreesAidedActive2015}. Recent attempts to aggregate a diverse set of participating units into simplified generic models have reduced the impact of varying latencies for frequency contingency services~\cite{fengProvisionContingencyFrequency2023}. Nevertheless, communication constraints continue to pose challenges for \glspl{VPP} in \gls{FFR} applications, requiring low latency (<500ms) while executing complex control algorithms that include non-transparent black-box models and robust optimization~\cite{brooksDemandDispatch2010}.
This is precisely where the recently developed concept of the power electronic-based system called \gls{ASG} can improve latencies and reduce complexity by acting as a physical aggregator of distributed resources~\cite{waldVirtualSynchronousMachine2024}. Effectively exploiting existing \gls{DG} units based on their local control objectives and responses without the need for additional communication infrastructure. Staying in the power plant analogy, the \gls{ASG} essentially operates an \textit{aggregated} power plant, instead of a \textit{virtual} power plant. A schematic illustration of this concept is presented in \cref{fig:asynch-concept} and explained as follows.
In this technology, an asynchronous grid connection is established by the AC/AC block in the center of the figure, representing any power electronic system capable of that type of connection, e.g., \textit{Three-Stage Solid-State Transformer}, \textit{AC-AC Direct Matrix Converter} or a \textit{Back-to-Back Converter} in series with a \gls{LFT}. Recent research has shown that the idea of creating an asynchronous cluster within a grid is technically feasible, as validated by experimental \gls{PHIL} testing~\cite{waldVirtualSynchronousMachine2024, waldAdaptiveVirtualSynchronous2023}. Detailed information on design parameters and power electronics associated with this technology are available in a previous study where a prototype was tested and validated~\cite{waldVirtualSynchronousMachine2024}.

This paper integrates a \gls{SST}-like system as \gls{FFR}-enhancing device within a power system and studies its system-level impacts using computer simulation. First, a system-level representation of the technology is developed using experimental measurements from \gls{PHIL} tests. Then, the obtained model is integrated into an IEEE 9-bus system to conduct dynamic performance analysis in DigSilent PowerFactory~\cite{sauerPowerSystemDynamics2017} . This analysis offers a proof of concept for the adoption of this new technology %and is sufficient to demonstrate the stability impact of y. 
and focuses on the assessment of the \gls{ASG}'s impact on power system dynamic performance. %. %In other words, the aim is to show that . %does not cause any harm to the stability and dynamic performance of the grid. 
To achieve this, the behavior of the system with the \glspl{FFR}-enhanced technology is then compared with that of the base case, which is simply the case without the \gls{ASG}, and the results indicate an improvement in the system's dynamic performance, proportional to the capacity of the employed \gls{ASG}.
Subsequently, the economic implications of such concepts are examined and how they can participate in the energy markets. To this end, the financial value of the \gls{FFR} service is evaluated, based on a specific market tariff and operational limits discovered during the technical assessment. For the market analysis, minor enhancements to the local control of \gls{DG} units are proposed that allow their aggregation by deploying an \gls{ASG} within the Eastern Interconnection/PJM. 
It is recognized that this enhancement introduces the mix of a North American tariff with a European connection rule for low voltage (LV) generators, and that these provisions have been designed according to the specific needs of their region. However, this amalgamation allows for the demonstration of how a different approach to \gls{FFR} tariffs or the enhancement of connection rules for \gls{DG} units could yield increased \gls{FFR} capabilities on both sides of the Atlantic, which is precisely one of the objectives of this study. 
Finally, the market analysis is used to form recommendations on the market environment and policy.

The contributions of this paper are threefold, and they are put forth as follows.

\begin{enumerate}
  \item This paper evaluates the dynamic performance enhancements of \gls{ASG} systems for \gls{FFR}, quantifying their contribution to power system stability and support.
  \item It develops a techno-economic analysis of \gls{ASG} systems for \gls{FFR}, building a business case by quantifying the value proposition and the revenue potential.
  \item It provides a critical perspective on market structure and incentives for \gls{DG}, focusing on \gls{SST} and regulatory considerations to hedge investment risk.
\end{enumerate}

In addition to the \gls{FFR} service provision explored in this study, the physical, asynchronous aggregation, and clustering of the power system sections can yield advantages such as: intentional islanding; enabling autonomous downstream grid operation; upstream grid support, e.g., reactive power compensation and limiting fault and disturbance propagation.

The remainder of this paper is organized as follows. 
\cref{sec:model} presents a detailed explanation of the model development of the \gls{ASG} system. \cref{sec:transient_ana} evaluates the impacts of this technology on the power system dynamic performance. \cref{sec:economic} develops a comprehensive analysis of the economic implications of the proposed technology. \cref{sec:disscusion} presents a critical opinion about market incentives and regulations that could help mitigate investment risk. \cref{sec:conclusion} concludes and closes the paper.

%% file: sections/02_modeling.tex
\section{Asynchronous Grid Response Modeling}\label{sec:model}

The recently developed asynchronous grid connection technology \cite{waldVirtualSynchronousMachine2024} utilizes two Back-to-Back connected low-voltage converters in series with a conventional \gls{LFT}. This configuration ensures sufficient power supply to a downstream grid, as shown in \cref{fig:asynch-concept}. 
\gls{PHIL} testing of the prototype, successfully demonstrated the control of the active power flow at device level around its original operating point. This control of active power flow allowed the support of the upstream grid during frequency contingency events.

\begin{figure*}[t]
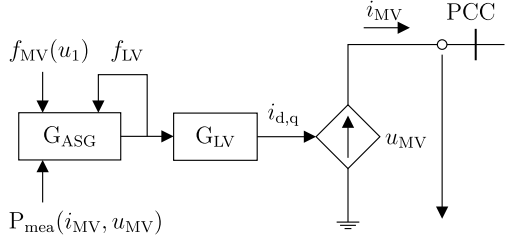

\centering
\include{img/pgf_tex_files/asg_blockdiagram_model}
\caption{\gls{ASG} model, including controller and converter average model, implemented within an RMS simulation in DigSilent Power Factory.\label{fig:model}}
\end{figure*}
The active power flow control concept is based on a form of \gls{DSR} that is mandatory for all converter-interfaced generation units newly deployed within the German power system due to the standards VDE-AR-N 4110 and 4105 for medium voltage (MV) and LV respectively~\cite{TechnicalRulesConnection2023,GeneratingSystemsLowvoltage2018}. The standard requires the implementation of a droop characteristic, essentially creating an active power-to-frequency sensitivity, defined as the variation of active power with respect to a frequency change, as follows:
\begin{equation}\label{eq:LoadSensi_P}
    \gls{Kpf}=\frac{\Delta P / P_0}{\Delta f / f_0}
\end{equation}
where $\gls{Kpf}$ is the sensitivity of the units' active power to frequency, $f$ is the frequency, $P$ is the active power, $f_0$ is the nominal frequency, and $P_0$ is the rated active power. Following the above-mentioned German standards, if the frequency of a LV grid is outside of a [49.8, 50.2]\,Hz deadband, the active power of a generation unit must follow the frequency with a sensitivity of $\gls{Kpf}=-20$ for the generation units and $\gls{Kpf}=-20$ (resp. $\gls{Kpf}=50$) for an energy storage unit for over-frequency (resp. under-frequency).
The asynchronous grid connection then utilizes the flexibility to adjust the downstream frequency ($f_{\text{LV}}$) within the boundaries of the grid code and therefore enforces an active power response from the downstream units. Using this mechanism, the active power flow can be controlled around the current operating point  which is an effective lever for the provision of a fast-frequency response service.

For the following system integration analysis, an abstracted model of the \gls{ASG} technology was developed that captures its non-linear properties and matches the behavior observed in \gls{PHIL} testing. Focusing on \gls{FFR} applications, the model represents active power behavior and control processes over several seconds while omitting less significant dynamics. The mathematical model development is detailed below and validated against experimental data to ensure accuracy.

\subsection{Mathematical Model}\label{sec:model_grid_code}

A reduced-order representation of the control system that operates the proposed \gls{ASG} technology is visualized by the block diagram shown in \cref{fig:model}. The governing principles of this control system and their mathematical models are explained throughout this subsection. 

A set of equations can adequately describe the \gls{ASG} model as follows: 

\begin{align}
    \gls{Ggov}(s) &\triangleq \gls{Pgov} =  (f_{\text{LV}}^*-f_{\text{LV}}) \left(\gls{Kpgov} + \frac{\gls{Kigov}}{s}\right)\label{eq:pi_gov}\\
    \gls{Gprop}(s) &\triangleq P_{\text{ff}} =\frac{1}{R}\cdot \Delta f_{\text{MV}}+ \gls{Pgov}\label{eq:f_prop}\\
    \gls{Gvsm}(s) &\triangleq \omega^* = \frac{1}{sT_\textrm{a}}\left(\gls{Pff} - P_{\text{mea}} - \frac{1}{D_\text{p}}\Delta\omega\right)\label{eq:vsm}
\end{align}
subject to:
\begin{equation}
    -\text{P}_{\text{lim}}\leq \text{P}_{\text{ff}} \leq \text{P}_{\text{lim}}\label{eq:plim}
\end{equation}

In \cref{eq:pi_gov}, \gls{Ggov}(s) represents the transfer function of the governor, which is a PI controller with the proportional gain of \gls{Kpgov} and the integral gain of \gls{Kigov}. 
Its goal is to steer the frequency of the downstream LV grid ($\Delta f_{\text{LV}}$) back to its nominal value by adjusting the active power set point (\gls{Pgov}). 
The active power set point \gls{Pgov} is subsequently altered in \cref{eq:pi_gov}, by a feed-forward summation of the amplified MV frequency deviation ($\Delta f_{\text{MV}}$) as in \cref{eq:f_prop}, where $R$ is the droop coefficient used to amplify that deviation. 
The downstream LV grid-forming controller uses a \gls{VSM} to regulate the phase angle of the supplied voltage. Multiple factors influenced the choice of the controller. Among others the ability to operate in grid-forming mode as well as grid-following and the flexibility to regain the nominal frequency value in a quick and smooth manner, lead to a combination of \gls{VSM} and PI governor controller.
The transfer function of the \gls{VSM} \cref{eq:vsm}, is based on a first-order swing equation proposed by Sakimoto et al. in 2011~\cite{k.sakimotoStabilizationPowerSystem2011}. 
It consists of the damping coefficient $D_\text{p}$ and has a time constant $T_\textrm{a}$ that represents the virtual moment of inertia. 
The input variables are the measured electric active power $P_{\text{mea}}$ and the enhanced active power set point $P_{\text{ff}}$, with the goal of adjusting the frequency set point $\omega^*$ to achieve the desired active power flow. 
The active power set point is subject to the limitation expressed in \cref{eq:plim}. 
This is needed to ensure the LV frequency set point $\omega^*$ stays within the LV distribution grid code boundaries, 47.5-51.5\,Hz~\cite{VoltageCharacteristicsPublic2020}.

The reaction of the active LV distribution grid to the frequency set point $\omega^*$ is based on a simple droop characteristic with coefficient $\gls{Kpf}(\gls{Kpf_scaling})$, shown in \cref{eq:glv}. This factor is subject to the composition of load/generation in the LV distribution grid at that specific time, defined by $\gls{Kpf_scaling}$, where $\gls{Kpf_scaling} \in [0, 1]$, representing 0-100\% of power absorbed/injected by active units obeying the German grid codes mentioned previously. 

\begin{equation}
\gls{Glv} \triangleq \gls{Pasg}(i_{\text{d,q}})= \begin{cases} 
\omega \cdot \gls{Kpf}(\gls{Kpf_scaling}), & \text{if } |\Delta\omega| \geq 200 \text{\,mHz}\\
0 & \text{otherwise}
\end{cases}\label{eq:glv}
\end{equation}
subject to:
\begin{equation}
0 \leq \gls{Kpf_scaling} \leq 1 
\end{equation}

The different control processes described above are highlighted by colors in the block diagram shown in \cref{fig:model}. The \gls{ASG} controller with the governor, frequency propagation, and swing equation (\crefrange{eq:pi_gov}{eq:vsm}) is located in the blue box. In the green shaded box, the calculation of the power response of the LV distribution grid droop characteristic (\cref{eq:glv}) takes place. The LV-side calculation is followed by the set point feed-through to the electrical representation of the \gls{ASG} and its connection to the power system in the yellow box. The transfer functions in \cref{fig:model} represent a grid-forming converter model based on a PI governor and a \gls{VSM}, with the addition of a power set point adaptation based on the upstream frequency. This adaptation leads to the active power adjustment of LV \gls{DG}, which is injected/absorbed by the current source in the yellow box.

As depicted in the yellow box, the upstream converter of the \gls{ASG} functions as a current source due to its grid-following operation with supporting features. Calculated currents are interfaced with the interconnection point to the grid. %Bus 5 of the IEEE 9-Bus System \cite{sauerPowerSystemDynamics2017}. 
Since active power control forms the foundation for \gls{FFR}, the developed model emulates the active power of the LV distribution network interfaced by an asynchronous grid connection.
\begin{figure}[t]
\centering
\includegraphics[width=\columnwidth]{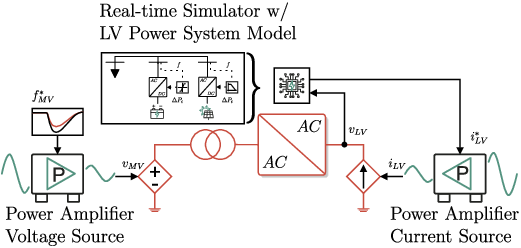}
\caption{Back-to-Back Converter in the double \gls{PHIL} test bed.\label{fig:phil}}
\end{figure}

\subsection{Model Validation - Power Hardware-in-the-Loop Testing}

To ensure accuracy of the derived model, it was validated against the experimental data collected in a previous study~\cite{waldVirtualSynchronousMachine2024}. The validation test case is based on a \gls{PHIL} experiment, testing the ability of an \gls{ASG} system to adjust the downstream active power set point using the frequency of voltage $v_\text{LV}$. A single line diagram of the setup is illustrated in \cref{fig:phil}, where the back-to-back converter is shown in the center, connected to two power amplifiers, representing the medium and low voltage power systems, respectively. The amplifiers are controlled by a central \textit{Real-time Simulator} executing both a single machine MV power system model and the LV distribution grid model, using a 24$\,\mu s$ time step. 
In \cref{fig:phil}, the MV power system is represented as a voltage source ($v_\textrm{MV}$), while the primary (left) side of the \gls{ASG} directly connected to the medium voltage side is acting as a grid-following active front-end converter, which can be model as a current source (analog to the yellow section in \cref{fig:model}). This current source behavior is responsible for supplying the secondary side LV converter with sufficient power. The secondary side converter is thus operating as grid-forming node.  It is supplying the LV distribution grid with the needed power and can be modeled as a voltage source connected to the emulated LV grid represented by a current source ($i_\textrm{LV}$).
\begin{figure}[t]
\begin{subfigure}[b]{\columnwidth}
    \centering
    \includegraphics{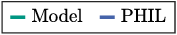}
\end{subfigure}
\vspace{0.1cm}
\begin{subfigure}[b]{\columnwidth}
    \centering
    \includegraphics[width=0.95\columnwidth]{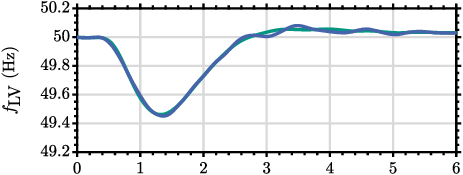}
    \caption{}
    \label{fig:f_comp_model_val}
\end{subfigure}
\vspace{0.1cm}
\begin{subfigure}[b]{\columnwidth}
    \centering
    \includegraphics[width=0.95\columnwidth]{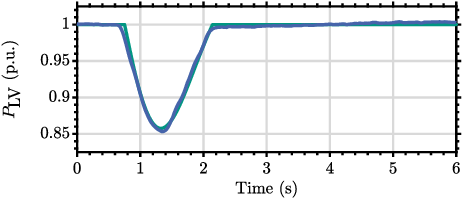}
    \caption{}
    \label{fig:p_comp_model_val}
\end{subfigure}
    \caption{The green line in a) shows the low-voltage frequency of the developed model, while the blue line represents experimental benchmark acquired in the \gls{PHIL} experiment. As a result of the frequency the active power flow adaptation is illustrated in b), using the same color code.\label{fig:model_val}}
\end{figure}
The parameters used for this validation are based on the parameters used in a previous hardware-related study. The \gls{RMS} deviation is used as the metric to quantify the precision of the analytical model and the data from the PHIL tests. The model fitting was conducted by minimizing the following objective function:
\begin{equation}
\text{RMSD} = \sqrt{\frac{1}{n} \sum_{i=1}^{n} \left( y_{\text{exp}}(t_i) - y_{\text{sim}}(t_i, \gls{para_vec}) \right)^2}\label{eq:obj_func}
\end{equation}
Here, $y_{\text{exp}}(t_i)$ represents the experimental data at time $t_i$, $y_{\text{sim}}(t_i, \gls{para_vec})$ represents the simulated data at time $t_i$ given the parameter vector \gls{para_vec}, and $n$ is the number of data points within the specified time window. The final parameters used are given in the appendix in \cref{tab:para_model}.

The test event is triggered by a frequency disturbance on the MV side of the connection ($f_\text{MV}$), which causes the system to initiate the \gls{FFR}. $f_{\text{LV}}$ in \cref{fig:f_comp_model_val}, represents the reaction of the downstream frequency of the developed model in green, and the actual \gls{PHIL} result in blue. The \gls{RMS} deviation between the model and experimental results yields a 0.027\,\% mean deviation during the frequency disturbance, which is small, thus, acceptable.
\cref{fig:p_comp_model_val}, illustrates the resulting adjustment of the active power flow due to the altered set point of the frequency $f_{\text{LV}}$. The RMS deviation between the active power curves yields a 0.5\,\% mean deviation, which is also small and acceptable. These results validate the mathematical model with sufficient accuracy to represent the proposed \gls{FFR} response in system-level integration studies, which will be the subject of investigation in the subsequent section.

%% file: img/pgf_tex_files/asg_blockdiagram_model.tex
% Styles (as you provided)
\tikzstyle{block} = [draw, rectangle, fill=gray!20,
    minimum height=1.8em, minimum width=2.5em, inner sep=2pt]
\tikzstyle{largeblock} = [draw, rectangle, 
    minimum height=2em, minimum width=3em, inner sep=2pt]
\tikzstyle{wideblock} = [draw, rectangle, 
    minimum height=1.8em, minimum width=4em, inner sep=2pt]
\tikzstyle{gain} = [scale=0.7, line width=0.6pt, draw, regular polygon, regular polygon sides=3, shape border rotate=180, minimum height=0.8em, minimum width=0.8em, inner sep=0pt, text height=1.5ex, text depth=0.25ex, fill=gray!20]
\tikzstyle{gain_right} = [scale=0.7, line width=0.6pt, draw, regular polygon, regular polygon sides=3, shape border rotate=270, minimum height=0.5em, minimum width=0.5em, inner sep=-2pt, text height=1ex, text depth=0.2ex, fill=gray!20]
\tikzstyle{gain_left} = [scale=0.7, line width=0.6pt, draw, regular polygon, regular polygon sides=3, shape border rotate=90, minimum height=0.5em, minimum width=0.5em, inner sep=-1pt, text height=1ex, text depth=0.2ex, fill=gray!20]

\tikzstyle{sum} = [draw, circle, node distance=1cm, minimum size=0.8em, inner sep=1pt]
\tikzstyle{input} = [coordinate]
\tikzstyle{output} = [coordinate]
\tikzstyle{connection} = [thick]

% Deadband block
\tikzstyle{deadband} = [
  draw,
  fill=gray!20,
  line width=0.3pt,
  inner sep=2pt,
  minimum size=8mm,
  path picture={
    \begin{scope}[line cap=round]
      \coordinate (SW) at (path picture bounding box.south west);
      \coordinate (SE) at (path picture bounding box.south east);
      \coordinate (NW) at (path picture bounding box.north west);
      \coordinate (NE) at (path picture bounding box.north east);
      \def\inset{0.12}
      \coordinate (ISW) at ($(SW)!\inset!(NE)$);
      \coordinate (ISE) at ($(SE)!\inset!(NW)$);
      \coordinate (INW) at ($(NW)!\inset!(SE)$);
      \coordinate (INE) at ($(NE)!\inset!(SW)$);
      \coordinate (IW) at ($(ISW)!0.5!(INW)$);
      \coordinate (IE) at ($(ISE)!0.5!(INE)$);
      \coordinate (IN) at ($(INW)!0.5!(INE)$);
      \coordinate (IS) at ($(ISW)!0.5!(ISE)$);
      \draw[->,gray!60] (IS) -- (IN);
      \draw[->,gray!60] (IW) -- (IE);
      \draw[very thick]
        (INW) -- ($(IW)!0.25!(IE)$) -- ($(IW)!0.75!(IE)$) -- (ISE);
    \end{scope}
  }
]

\tikzstyle{lim} = [
  draw,
  fill=gray!20,
  line width=0.3pt,
  inner sep=2pt,
  minimum size=8mm,
  path picture={
    \begin{scope}[line cap=round]
      % corners of the node's bounding box
      \coordinate (SW) at (path picture bounding box.south west);
      \coordinate (SE) at (path picture bounding box.south east);
      \coordinate (NW) at (path picture bounding box.north west);
      \coordinate (NE) at (path picture bounding box.north east);
      % inset: fractional margin inside the box for the internal drawing
      \def\inset{0.12}
      \coordinate (ISW) at ($(SW)!\inset!(NE)$);
      \coordinate (ISE) at ($(SE)!\inset!(NW)$);
      \coordinate (INW) at ($(NW)!\inset!(SE)$);
      \coordinate (INE) at ($(NE)!\inset!(SW)$);
      % limiter curve: bottom plateau -> rising slope -> top plateau
      % tune the 0.xx values to change plateau lengths/turn-on point
      \coordinate (P1) at ($(ISW)!0.12!(ISE)$);  % start of bottom plateau
      \coordinate (P2) at ($(ISW)!0.38!(ISE)$);  % end of bottom plateau
      \coordinate (P3) at ($(INW)!0.78!(INE)$);  % end of rising slope
      \coordinate (P4) at ($(INW)!0.94!(INE)$);  % top plateau near right
      \draw[very thick] (P1) -- (P2) -- (P3) -- (P4);
    \end{scope}
  }
]

% Minimalistic group-box style (used with fit)
\tikzstyle{groupbox} = [draw, rounded corners, line width=0.6pt, inner sep=7pt]

\begin{tikzpicture}[auto, node distance=1.5cm,>=latex']

  % Frequency/Active Power Control Loop
  \node [input, name=origin] {};
  \node [sum, right of= origin, node distance=0.65cm] (sum_omega) {};
  \node [block, right of=sum_omega, node distance=1.25cm] (pi_omega) {PI};    
  \node [sum, right of=pi_omega, node distance=1.25cm] (sum_p_fp) {};
  \node [lim, right of = sum_p_fp, node distance=1cm] (p_fp_lim) {};
  \node [gain_left, below of =p_fp_lim, node distance=2.75cm] (omega_gain) {$1/2\pi$};
  \node [sum, right of=p_fp_lim, node distance=1.25cm] (sum_p) {};
  \node [gain,  above of=sum_p_fp, node distance=2cm] (K_fp) {$1/R$};

  % Swing equation block
  \node [block, right of=sum_p, node distance=1.25cm] (inertia) {$\frac{1}{sT_\textrm{a}}$};
  \node [sum, right of= inertia, node distance=1.75cm] (sum_omega_out) {};
  \node [deadband, right of = sum_omega_out, node distance=1.9cm] (deadband) {};
  \node [gain_right, right of = deadband, node distance = 2cm] (Kpf_gain) {$\textrm{K}_\textrm{pf}(\alpha)$};

  \node [above of = deadband, node distance= 0.75cm] (){\small Deadband};

  % Damping feedback
  \node [block, below of=inertia, node distance=1.5cm] (damping) {$\frac{1}{D_\textrm{p}}$};

  % Reference inputs
  \node [input, above of=sum_omega, node distance=1cm] (omega_ref_in) {};
  \node [input, above of=sum_omega_out, node distance=1cm] (omega_0_in) {};
  \node [input, above of=sum_p, node distance=1cm] (P_ref_in) {};
  \node [input, above of=K_fp, node distance=1cm] (P_fp_ref_in) {};
  % Feedback signals
  \node [input, below of=sum_omega, node distance=1cm] (omega_fb_in) {};

  % % Connections - Frequency/Active Power Loop
  % \draw [connection,-stealth] (omega_in) -- node[above, pos=0.8] {$-$} node[above, near start] {$f_\textrm{LV}$} (sum_omega);
  \draw [connection,-stealth] (origin) -- node[near start, above] {$f_\textrm{LV}^*$} (sum_omega);
  \draw [connection,-stealth] (sum_omega) -- (pi_omega);
  \draw [connection,-stealth] (pi_omega) -- node[above] {$P_\textrm{gov}$} (sum_p_fp);
  \draw [connection,-stealth] (sum_p_fp) -- node[above] {} (p_fp_lim);
  \draw [connection,-stealth] (p_fp_lim) -- node[above] {$P_\textrm{ff}$} (sum_p);
  \draw [connection,-stealth] (P_ref_in) -- node[right, pos=0.8] {$-$} node[near start, left] {$P_{\text{mea}}$} (sum_p);
  \draw [connection,-stealth] (P_fp_ref_in) -- node[near start, left] {$\Delta f_\textrm{MV}$} (K_fp);
  \draw [connection,-stealth] (K_fp) -- (sum_p_fp);
  \draw [connection,-stealth] (sum_p) -- (inertia);
  \draw [connection,-stealth] (inertia) -- node[above] {$\Delta\omega^*$} (sum_omega_out);
  \draw [connection,-stealth] (omega_0_in) -- node[near start, left] {$\omega_0$} (sum_omega_out);
  \draw [connection,-stealth] (sum_omega_out) -- node[above] {$\omega^*$} (deadband.west);
  \draw [connection,-stealth] (deadband.east) -- (Kpf_gain.west);

  % Damping feedback connections
  \draw [connection,-stealth] ($(inertia)!0.65!(sum_omega_out)$) |- (damping);
  \draw [connection,-stealth] (damping) -| node[very near end, left] {$-$} (sum_p);

  % Frequency feedback

  \draw [connection,-stealth] ($(sum_omega_out)!0.2!(deadband.west)$) |- (omega_gain) -| node[left, pos=0.85] {$f_\textrm{LV}$} node[left,pos=0.95] {$-$}(sum_omega);

  % ===== Circuitikz part anchored to Kpf_gain output and the current source =====
  % Put the current source right next to Kpf_gain and feed it from Kpf_gain output
  % Center of the current source relative to Kpf_gain.east
  \coordinate (CSctr) at ($(Kpf_gain.east)+(2,0)$);
  \coordinate (CSTop) at ($(CSctr)+(0,0.9)$);
  \coordinate (CSBot) at ($(CSctr)+(0,-0.9)$);
  % Draw the (controlled) current source
  \draw (CSBot) to[cI, name=cSource] (CSTop);

  \draw (CSBot) node[tlground] (gnd) {};
  % Feed from Kpf_gain to the source (into its center)
  \draw[connection,-stealth] (Kpf_gain.east) -- node[above] {$P_\textrm{ASG}$} (cSource);

  % Now place the rest of the circuit RELATIVE to the current source (same structure as your commented block)

  % Rightmost open terminals for voltage measurement
  \coordinate (Terminaltop)  at ($(CSTop)+(1.3,0)$);
  \coordinate (Terminalbase) at ($(CSBot)+(1.3,0)$);
  \coordinate (Bus5) at ($(Terminaltop)+(0.5,0)$);
  \node[ocirc] (TerminaltopNode) at (Terminaltop) {};

  \draw (Bus5) \bushere{0.4}{PCC}{} ;

  % Measurement open (vertical)
  \draw (Terminalbase) to[open, v_=$ $, straight voltages, name=VDC] (Terminaltop);
  % Labels (optional; adjust or comment out if you don't use glossaries)
  \node at ($(VDC)+(0.4,0)$) {$v_\textrm{MV}$};

  % Horizontal stubs connecting the branches
  \draw (CSTop) to[short, f^>=$i_\textrm{MV}$] (TerminaltopNode.west);
  \draw (TerminaltopNode.east) to[short] (Bus5) to[short] ++(0.5,0) coordinate(PCC);

  % -------- Group boxes (background) --------
  \begin{scope}[on background layer]

       % Circuit box with manual top padding (to avoid clipping labels)
    % Adjust these two paddings to taste:
    \def\circTopPad{0.75cm}   % increase to raise top of circuit box
    \def\circBotPad{0.5cm}   % increase to lower bottom of circuit box
    % Phantom coordinates to enlarge the circuit box vertically
    \coordinate (circTopPadC) at ($(Terminaltop)+(0,\circTopPad)$);
    \coordinate (circBotPadC) at ($(gnd)+(0,-\circBotPad)$);
    \node[groupbox, fill=kit-yellow!10,
      fit=(cSource)(TerminaltopNode)(PCC)(gnd)(Terminalbase)(circTopPadC)(circBotPadC)] (box_circuit) {};

    % LV grid model (deadband + Kpf gain), aligned to circuit box height
    % Project the circuit's top/bottom onto the LV group's horizontal position
    \def\circLeftPad{0.22cm}
    \coordinate (lv_top) at (deadband |- circTopPadC);
    \coordinate (lv_bot) at (deadband |- circBotPadC);
    \coordinate (leftPad) at ($(deadband.west)-(\circLeftPad,0)$);
    
    \node[groupbox, fill=kit-green!10,
      fit=(leftPad)(deadband)(Kpf_gain)(lv_top)(lv_bot)] (box_lv) {};

    \coordinate (rightPad_active) at ($(sum_omega_out)!0.2!(deadband.west)$);
    % Active power loop
    \node[groupbox, fill=kit-blue!10,
      fit=(origin)(sum_omega)(pi_omega)(sum_p_fp)(sum_p)(K_fp)(inertia)(sum_omega_out)(damping)(P_fp_ref_in)(omega_gain) (rightPad_active)
      ] (box_active) {};

  \end{scope}

\end{tikzpicture}

%% file: sections/03_transient_analysis.tex
\section{System Integration and Dynamic Analysis}\label{sec:transient_ana}

This section seeks to understand the fast-frequency response capability that an asynchronous grid connection can offer and its impact on the dynamic performance and system stability when integrated within a power system. 
The dynamic performance analysis was conducted using the modified IEEE 9-bus benchmark power system~\cite{vijayvittalPowerSystemControl2020, sauerPowerSystemDynamics2017}. 

\begin{figure}[t]
\centering
\input{img/pgf_tex_files/ninebus.tex}
\caption{Single-line diagram of the IEEE 9-bus system~\cite{vijayvittalPowerSystemControl2020, sauerPowerSystemDynamics2017}.\label{fig:ninebus}}
\end{figure}

\subsection{Computer Simulation Procedure}

Implementation in DigSilent Power Factory began with initial parameter validation of the \gls{ASG}. A \textit{Static Generator} block implements the response algorithm, acting as a controllable power source capable of injecting or absorbing active power calculated by the \gls{ASG} control structure shown in \cref{fig:model}.
Through an automated Python interface, the combined IEEE 9-bus system and \gls{ASG} model enables iterative parameter changes and comparative analysis across different cases. \cref{fig:ninebus} illustrates the single-line diagram of the 9-bus system, including the \gls{ASG} and the load step used as a frequency disturbance trigger. The detailed parameters are provided in tabular form in the Appendix in \cref{tab:ninebus_sys_para}. The controllable power source/sink block representing the \gls{ASG} responds to system frequency dynamics by injecting or absorbing power at bus 5 (the load bus between generator 1 and generator 2). Bus 5 was chosen arbitrarily after a preliminary assessment revealed that the \gls{ASG} location in this system was not a determining factor for its impact on the system's frequency response.

The analysis involved four distinct scenarios: a base case without an \gls{ASG} system, and three configurations with \gls{ASG} systems rated at 2\,MW, 5\,MW, and 10\,MW respectively. Each scenario faced disturbance through a +5\,\% load change at Bus 6, equivalent to 4.5\,MW additional load power (red shaded step block in \cref{fig:ninebus}). While this creates a relatively large frequency deviation exceeding 250\,mHz, it effectively demonstrates the proposed concept's capabilities.

The resulting time series of the frequency at each bus is subsequently transferred to MATLAB using the Python library \textit{MATLAB Engine}. For time-domain analysis and the subsequent economic evaluation, MATLAB and Python serve as the main tools.

\subsection{Time-Domain Analysis}
\begin{figure}[t]
    \begin{subfigure}[b]{\columnwidth}
        \centering
        \includegraphics{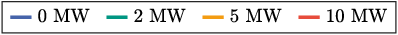}
    \end{subfigure}
    
    \vspace{0.1cm}
    \begin{subfigure}[b]{\columnwidth}
        \centering
        \includegraphics[width=0.98\columnwidth]{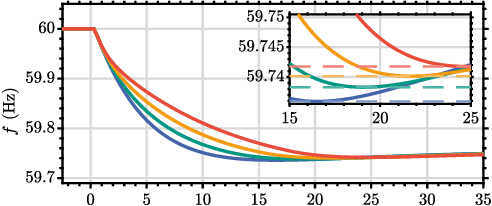}
        \caption{}
        \label{fig:f_response}
    \end{subfigure}
    \vspace{0.3cm}
    \begin{subfigure}[b]{\columnwidth}
        \centering
        \includegraphics[width=0.98\columnwidth]{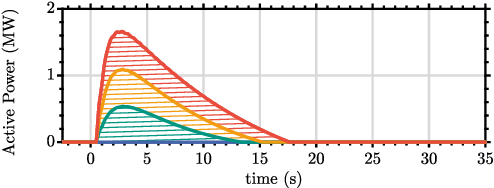}
        \caption{}
        \label{fig:p_response}
    \end{subfigure}
    \caption{Plot of the system frequency at bus 6 in a) for a frequency response of the \gls{ASG} for different power ratings of the \gls{ASG}, the respective active power flow shown in b).}
    \label{fig:comp_power_var}
\end{figure}

\begin{table}[t]
    \setlength{\tabcolsep}{4pt}%
    \renewcommand{\arraystretch}{1.1}%
    \centering
    \caption{Comparative performance of different cases equipped with \gls{ASG} with respect to the base case - the measurements are from bus 6.}
    \begin{tabular}{r|cc|cc|c}
        % Sum = 7.5cm
        \toprule
        $P_{\gls{ASG}}$ &  $\Delta \mathrm{t}_{\mathrm{NADIR}}$ & $\%$ to Base & RoCoF & $\%$ to Base & \multicolumn{1}{c}{$\Delta$E (kWh)}\\
        \midrule
        \color{kit-blue}Base   & 16.44\,s & 0.00   &5.2 $\times 10^{-2}$ & 0.0 &0.00 \\
        \color{kit-green}2\,MW   & 19.20\,s & +15.78    &5.0 $\times 10^{-2}$ & -5.7 &1.95\\
        \color{kit-yellow}5\,MW    & 21.60\,s & +31.38  &4.6 $\times 10^{-2}$ & -12.5 &10.99\\
        \color{kit-red}10\,MW       & 24.78\,s & +50.70 &4.2 $\times 10^{-2}$ & -20.4 &38.15\\
        \bottomrule
    \end{tabular}
    \label{tab:Test_Cases_Overview}
\end{table}

Upon system disturbance, the frequency drops due to a mismatch between generation and consumption. As shown in \cref{fig:comp_power_var}, the resulting frequency responses vary with different levels of \gls{ASG} participation, with the baseline behavior (without \gls{ASG}) shown in blue. Baseline measurements record the frequency nadir at 16.44\,s, while the \gls{ASG} implementation delays this occurrence. Systems equipped with 2\,MW, 5\,MW, and 10\,MW \gls{ASG} reach the nadir at 19.20\,s, 21.60\,s, and 24.78\,s respectively. Additionally, nadir frequency values improve from 59.736\,Hz (baseline) to 59.738\,Hz, 59.740\,Hz, and 59.742\,Hz with increasing \gls{ASG} ratings. As evident here, and as the core functionality of the \gls{ASG} suggests, its introduction aids the frequency response in the form of a delayed nadir time and marginally less frequency deviation (that is lower nadir). In addition, the \gls{RoCoF} is significantly reduced by 5.7\,\%, 12.5\,\% and 20.4\,\%. This frequency response support is safely attributed to the additional active power provided by the \gls{ASG}, as it is the only parameter that distinguishes these cases. For a more convenient depiction of this effect, the nadir section is magnified in the top right of \cref{fig:comp_power_var}. 

For comparative purposes, the response from the three cases with \gls{ASG} with respect to the base case was quantified using the time it takes for the frequency to reach its nadir in seconds, $\Delta \mathrm{t}_{\mathrm{NADIR}}$, and its percentage change (\%.) In addition, the amount of energy injected by \gls{ASG}, $\Delta$E was included. Which is the integral value of active power injection $\Delta$P, illustrated in \cref{fig:p_response} and highlighting $\Delta$E as the shaded area below each line. The quantitative results are shown in Table \ref{tab:Test_Cases_Overview}. Comparing the time from the start of the disturbance until the nadir is reached shows that a 2\,MW support system would increase the time by 15\,\%, with the possibility of nearly 50\,\% for a 10\,MW support system.

\subsection{Modal Analysis}

Looking at the time-domain traces, one could reasonably argue that all cases studied have similar frequency trajectories. Herein lies the motivation for modal analysis to learn how this technology may alter dynamical modes that are excited during the frequency response as the focal point of the analysis. To this end, Prony's analysis is applied to the time-domain frequency signals recorded in the simulation~\cite{sajadiSmallSignalStabilityAnalysis2019}. Prony's method is a data-driven technique for studying dynamical systems because of its ability to estimate eigenvalues of a nonlinear system that emerge from small-signal disturbances by using a measured data sequence to fit a linear combination of complex exponential terms with damping in form of \cite{fernandez2018coding}: 
\begin{equation}
    f(t) = \sum_{i=1}^{N} A_i  e^{\sigma_i t}  \text{cos} (\omega_i t + \phi_i)  
\end{equation}
where $N$ is the number of terms and 
$A_i$ is the amplitude,
$\sigma_i$ is the damping factor, 
$\omega_i$ is the frequency, and
$\phi_i$ is the phase of $i$th term, representing the eigenvalue in the form of $\lambda_i=\sigma_i+\pm \boldsymbol{j} \omega_i$.

In this analysis, the two-term solution was sought; yielding a pair of complex conjugate roots, given that the frequency response of generators in a power system around its fundamental frequency resembles the response of second-order harmonic oscillators \cite{sajadi2022synchronization}. The outcome of the modal estimation regarding the frequency response for differing levels of support from the \gls{ASG} are shown in \cref{tab:pronys}.

\begin{table}[h]
\centering
\caption{Dominant modes of the frequency response.}
\label{tab:pronys} %%% From Gen 1
\begin{tabular}{rccc}
\toprule
\multicolumn{1}{c}{$\mathrm{P}_{\mathrm{\gls{ASG}}}$}  & $\lambda_{12}$ & $f$ & $\zeta$ \\  \midrule
\color{kit-blue}Base & -0.174 $\pm$ $\boldsymbol{j}$ 0.184  & 0.029  &   68.61    \\ 
\color{kit-green}2\,MW & -0.178 $\pm$ $\boldsymbol{j}$ 0.162  & 0.026 & 74.01 \\
\color{kit-yellow}5\,MW & -0.182 $\pm$ $\boldsymbol{j}$ 0.145 & 0.023 & 78.31 \\  
\color{kit-red}10\,MW & -0.186 $\pm$ $\boldsymbol{j}$ 0.127  & 0.020  &  82.55 \\ 
\bottomrule
\end{tabular}
\end{table}

The results yield a negative root for the oscillatory modes, which was expected for time-domain signals being all stable, in all cases considered. They show that the frequency of excited modes, denoted by f, becomes smaller proportional to the increase in contribution from the \gls{ASG}. This frequency is computed as $f_\mathrm{i}=(2\pi)^{-1} \omega_\mathrm{i}$. The lower this frequency, the less oscillatory the response; therefore, the more desired. The higher contribution from the \gls{ASG} technology also appears to increase the damping ratio of oscillatory modes, $\zeta$. The damping ratio is a dimensionless metric that relates to the decay of system oscillations following a transient. It is computed as $\zeta_\mathrm{i}=\frac{\sigma_\mathrm{i}}{\sqrt{\sigma_\mathrm{i}^2+\omega_\mathrm{i}^2}}\times 100$ and ranges between 0 and 100; the higher the damping ratio, the better the dissipation of transients. This suggests that the additional fast power injection by the \gls{ASG} technology helps eigenvalues migrate further away from the imaginary axis and simultaneously move closer to the real axis. The injection provides an additional damping-like support without changes to inertial momentum in the system. This observation is corroborated by the analysis shown in \cite{sajadi2022synchronization} for the signature characteristics of increased damping in the system.

The time-domain results and eigenvalue analysis collectively established that the use of the \gls{ASG} technology can offer a positive contribution by improving frequency nadir and ROCOF. Next, the financial viability of deploying such technology under current market incentives, tariff structures, and subject to performance limitations, will be explored.

%% file: img/pgf_tex_files/ninebus.tex
% \newcommand{\acachere}[3]{
%     coordinate(tmp)
%     node[rectangle]

%     (tmp)
% }

% Variant 2: with a light gray square around it (block-like)
\tikzstyle{stepblock} = [
  draw,
  fill=kit-red!20,
  line width=0.4pt,
  inner sep=2pt,
  minimum size=8mm,
  path picture={
    \begin{scope}[line cap=round]
      \def\inset{0.12}
      \coordinate (SW) at (path picture bounding box.south west);
      \coordinate (SE) at (path picture bounding box.south east);
      \coordinate (NW) at (path picture bounding box.north west);
      \coordinate (NE) at (path picture bounding box.north east);
      \coordinate (BL) at ($(SW)!\inset!(NE)$);
      \coordinate (BR) at ($(SE)!\inset!(NW)$);
      \coordinate (TL) at ($(NW)!\inset!(SE)$);
      \coordinate (TR) at ($(NE)!\inset!(SW)$);
      \coordinate (BC) at ($(BL)!0.5!(BR)$);
      \coordinate (TC) at ($(TL)!0.5!(TR)$);
      \draw[very thick] (BL) -- (BC) -- (TC) -- (TR);
    \end{scope}
  }
]

\tikzset{
  connection/.style={semithick},    
  % optional: define a default arrow tip to use with ->
  >={Stealth[length=2.2mm,width=2mm]}
}

\begin{tikzpicture}[scale = 0.5, american,>=stealth]

    \draw (0,0)node[oscillator, scale =0.8](2){} -- ++(1,0)
    \bushere{0.3}{Bus 2}{} coordinate(Bus 2);
    \draw (2.n) node[above] {$G_2$};
    \draw (Bus 2) to[oosourcetrans, name=trafo, l_=$T_2$] ++(2.5,0) \bushere{0.3*2}{Bus 7}{} coordinate(Bus 7);
    \draw (Bus 7) -- ++(2.5,0) \bushere{0.3*3}{Bus 8}{} coordinate(Bus 8);
    \draw (Bus 8) -- ++(2.5,0) \bushere{0.3*2}{Bus 9}{} coordinate(Bus 9);
    \draw (Bus 9) to[oosourcetrans, name=trafo, l_=$T_3$] ++(3,0) \bushere{0.3*1}{Bus 3}{} coordinate(Bus 3) ++(3,0) node[oscillator, scale =0.8](3){};
    \draw (Bus 3) -- ++(1,0) to (3.w);
    \draw (3.n) node[above] {$G_3$};
    \draw (Bus 7) ++(-1.5,-3) \busherehor{3}{}{Bus 5} coordinate(Bus 5);
    \draw (Bus 7) ++(0,-1) -| ++(-0.5,-2) ++(1,0) -- ++(0, -3) ++(2,0) \busherehor{4}{Bus 4}{} coordinate(Bus 4);
    \draw (Bus 4) ++(3.5,0) -- ++(0, 3) ++(1,0) \busherehor{2}{Bus 6}{} coordinate(Bus 6) ++(-1,0) |- ++(-0.5,2);
    \draw (Bus 4)  to[oosourcetrans, name=trafo, l_=$T_1$] ++(0,-3) \busherehor{1}{Bus 1}{} coordinate(Bus 1);
    \draw (Bus 1) ++(0.8,-2) node[oscillator, scale =0.8](1){};
    \draw (Bus 1) -- ++(0,-1) to (1.n);
    \draw (1.south) node[below] {$G_1$};

    %% Loads
    \draw (Bus 5) -- ++(0,-1) node[isosceles triangle,
	isosceles triangle apex angle=60,
	draw,
	rotate=270,
	fill=black,
	minimum size =0.3cm,anchor=west]{}  coordinate(Load C);
    \draw (Load C.north) ++(1.25,0) node[below]  {$L_C$};
    \draw (Bus 6) ++(0.5,0) -- ++(0,-1) node[isosceles triangle,
	isosceles triangle apex angle=60,
	draw,
	rotate=270,
	fill=black,
	minimum size =0.3cm,anchor=west]{} coordinate(Load B);
    \draw (Load B.north) ++(-1,0) node[below]  {$L_B$};
    \draw (Load B) ++(1.6,0)  node[stepblock, minimum size=5mm, anchor=north] (stepB) {} node[below right, xshift=0.2cm] {+5\,\%};
    \draw[connection, -Stealth] (stepB) -- ++(-1.3,0);

    \draw (Bus 8) ++(0,-1) -- ++(1,0) node[isosceles triangle,
	isosceles triangle apex angle=60,
	draw,
	fill=black,
	minimum size =0.3cm,anchor=west]{} coordinate(Load A);
    \draw (Load A.south) ++(0.5,-0.5) node[below]  {$L_A$};

    %% AG Support
    \draw(Bus 5) ++(-2,-2) node[twoportshape, fill=kit-green!20 ,t=ASG](){} coordinate(AG);
    \draw(Bus 5) ++(-2,0) -- ++(0,-1);
    % node[isosceles triangle,
	% isosceles triangle apex angle=60,
	% draw,
	% rotate=270,
	% fill=kit-green,
	% minimum size =0.5cm,anchor=west]{} coordinate(Load 0);
 %    \draw[kit-green] (Load 0.north) ++(1.25,0) node[below]  {$L_0$};

\end{tikzpicture}

%% file: sections/04_economics.tex
\section{Economic Evaluation}\label{sec:economic}

Having established technical feasibility, this section investigates the market participation potential of the \gls{ASG} technology in frequency response services, focusing on revenue stream identification. This analysis explores the deployment of the \gls{ASG} in North America, specifically within the Pennsylvania-New Jersey-Maryland (PJM) market. As a \gls{RTO} in the Eastern Interconnection of the United States, PJM operates an electric transmission system across 13 states within the Eastern Interconnection grid. This analysis proposes an enhancement to the PJM protocols related to the \gls{DG} units, following German standards, such that newly connected \gls{DG} units are required to provide active frequency response through droop characteristics, as discussed in \cref{sec:model} and described in \cref{eq:glv}~\cite{TechnicalRulesConnection2023, GeneratingSystemsLowvoltage2018}.

The proposed integration of PJM-style regulation markets with German-inspired \gls{DG} features would enable power electronic systems like the \gls{ASG} to aggregate \gls{DG} fleets physically, offering profitable frequency response services. Most notably, such an approach eliminates the need for additional communication infrastructure or complex control hierarchies typically required for aggregating dispersed \gls{DG}.

\begin{figure}[t]
    \begin{subfigure}[b]{\columnwidth}
        \centering
            \includegraphics{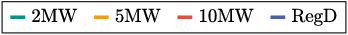}
    \end{subfigure}
    
    \vspace{0.1cm}
    
    \begin{subfigure}[b]{\columnwidth}
        \raggedright
        \includegraphics{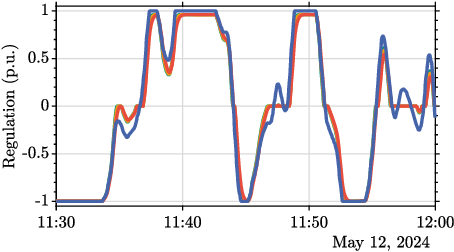}
        \caption{\label{fig:PJM_timeseries}}
    \end{subfigure}
    
    \begin{subfigure}[b]{\columnwidth}
        \raggedright
        \includegraphics{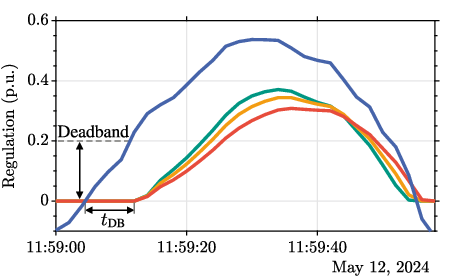}
        \caption{\label{fig:PJM_timeseries_30sec}}
    \end{subfigure}
\caption{The time series in a) shows 30 minutes of the regulation signal \textit{RegD} in blue paired with the \gls{ASG} regulation for different power ratings in green, yellow and red. b) is a zoomed in version of the same data, highlighting the difference between the power ratings and the impact of the deadband and ramp rate limitation. \label{fig:timeseries}}
\end{figure}

\subsection{PJM Regulation Concept - RegD/A}
PJM, operates a real-time ancillary service market, in which the frequency regulation service is based on an \gls{AGC} concept, sending regulation signals to participating units, to help maintain a balanced system. The signals for frequency regulation are divided into two signals, \textit{RegA} and \textit{RegD}. \textit{RegA} represents a low-pass filtered \gls{ACE} signal for traditionally slow-acting resources. \textit{RegD} is a high-pass filtered \gls{ACE} signal designed for fast response resources, e.g. \glspl{SST} or Energy Storage Systems.
In this scenario, given the fast-responding nature of the \gls{ASG}, the participation in the \textit{RegD} regulation scheme is investigated. PJM sends the signal in intervals of 2 seconds to participating systems, with normalized values of -1 to 1.
This value is not directly applicable as an active power set point within the \gls{ASG} controls; therefore, by rescaling it as outlined in \cref{eq:rescale} a scaled frequency set point will be generated.
\begin{align}
        f_{LVsc}^* &= \text{rescale(}regD\text{)}\cdot f_{LV}^*\label{eq:rescale}\\
        \gls{Ggov}(s) \triangleq \gls{Pgov} &=  (f_{LVsc}^*-f_{LV}) \left(\gls{Kpgov} + \frac{\gls{Kigov}}{s}\right)\label{eq:pi_gov_pjm}
\end{align}
In \cref{eq:rescale} the correct adjustment of the frequency set point is ensured. This in turn provokes the respective droop control reaction within the \gls{DG} units, due to the enhanced behavior introduced by the German standard VDE-AR-N 4105~\cite{GeneratingSystemsLowvoltage2018}. To adjust $f_{\text{LV}}^*$, which is in per unit and is usually set to 1, the \textit{RegD} signal is rescaled to adjust the frequency to up to $\pm$ 1\,Hz. For a 60\,Hz system that requires a rescaling of the \textit{RegD} signal from \(\left[-1\text{ }..\text{ }1\right]\) to \(\left[0.935\text{ }..\text{ }1.0165\right]\) as in \cref{eq:rescale}. The governor previously described by \cref{eq:pi_gov}, now \cref{eq:pi_gov_pjm}, uses the value $f_{\text{LVsc}}^*$ as the new set point of the nominal frequency.  

A 30-minute window of the simulation data is displayed in \cref{fig:PJM_timeseries}, the active power adjustment is only triggered after the frequency set point is outside the 200\,mHz deadband, which translates to 20\,\% of the allowed $\pm$ 1\,Hz frequency deviation. The influence of the deadband is illustrated in the zoomed plot, in \cref{fig:PJM_timeseries_30sec}, where the regulation signals react after the \textit{RegD} signal is outside the deadband of 20\,\%. Additionally, this plot highlights that a power electronic system with a higher power rating might be limited by the ramp rate at which it is allowed to change its output power, thus not reaching its nominal power before the regulation decreases its demand again. This leads to a reduced overall regulation, with respect to the power rating, which will subsequently impact the potential revenue, discussed below.

\subsection{Short-Term Revenue Forecast}

\begin{figure}[t]
    \centering
    \includegraphics[width=0.95\columnwidth]{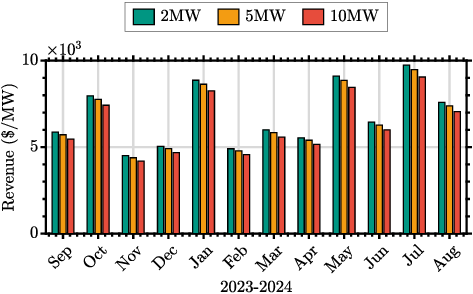}
    \caption{Monthly revenue within PJM regulation from 09/23-08/24 for three \gls{ASG} power ratings.}
    \label{fig:revenue_monthly}
\end{figure}

PJM's frequency response service compensation follows a pay-for-performance model, aligned with FERC Order No. 755. Compensation depends not only on the volume of regulated active power but also on signal adherence precision~\cite{FinalRuleOrder2011}. The performance evaluation combines actual integrated power regulation (capability or mileage, \gls{MRegD}) with performance score ($\rho$) to determine five-minute remuneration credits. A publicly available PJM excel tool yielded $\rho = 0.5$, resulting in a smaller value than the suggested default for demand response units ($\rho=0.76$)~\cite{PJMAncillaryServices}. In addition to capability and performance score, the mileage ratio $\beta_t^M$ is used and defined as:
\begin{equation}
    \gls{Mratio} = \frac{\gls{MRegD}}{\gls{MRegA}}\label{eq:m_ratio}
\end{equation}
where the mileage (\gls{MRegD}) is the integrated movement of the regulation control signal, as shown for \textit{RegD} below:
\begin{equation}\label{eq:mileage}
   \gls{MRegD} = \sum_{i =1}^{N}\vert \gls{RegD}_{i}-\gls{RegD}_{i-1}\vert    
\end{equation}
here N=150 is the number of 2-second samples in the 5-minute evaluation window.
This calculation process subsequently yields the regulation credit (\gls{Creg}) for a 5 minute window, based on the following equations: 
\begin{equation}
    \gls{Creg} = \gls{MRegD} \cdot \rho  \cdot( \gls{RCCP} +  \gls{Mratio} \cdot\gls{RPCP})\label{eq:credit}
\end{equation}
with the \textit{Regulation Capability Clearing Price} as \gls{RCCP} (in \$/\,MW), \gls{MRegD} (in MW) integrated regulation within the time window and the \textit{Regulation Performance Clearing Price} represented by \gls{RPCP} (in \$/$\Delta$\,MW). Both clearing prices are sourced from the historical data during the investigated time period. The database is freely accessible through PJM's open data miner interface and provides the full 5 minute resolution clearing prices~\cite{PJMDataMiner,PJMAncillaryServices}. During the study period (09/23-08/24), \gls{RCCP} ranged from 0 to 238.5\$/\,MW and \gls{RPCP} ranged from 0 to 2.8\$/$\Delta$\,MW, with average values of 2.2\$/\,MW and 0.08\$/$\Delta$\,MW respectively. In the scenario studied, an average distributed power generation of 2, 5 and 10\,MW within the downstream grid is assumed, offering its active power through the droop characteristic of \cref{eq:glv}. Using \cref{eq:credit}, the monthly revenue of an \gls{ASG} system indirectly controlling downstream \gls{DG}, is in the range of 4,000 - 10,000 \$/\,MW, as illustrated in \cref{fig:revenue_monthly}. The revenue bar chart also highlights the slightly reduced revenue per MW, for a system with higher power rating due to the ramp rate limitation and subsequently a worse performance score ($\rho$), as mentioned above.

\subsection{Long-Term Cash-Flow Forecast}
The long-term financial evaluation of the new technology begins with calculating the \gls{NPV}, which serves as a primary indicator of investment profitability over the chosen 15-year horizon. The \gls{NPV} uses the future cash flow (\gls{Cashflow}) discounting its value back to the present day to ensure that all future earnings and expenses are accounted for, the calculation is based on the following formula:
\begin{equation}
    \gls{NPV} = \sum_{t=0}^{T} \frac{\gls{Cashflow}(t)}{(1 + \gls{interest})^t}
\end{equation}
where T is the total number of periods (years) and \gls{interest} is the discount rate applied to the cash flows. The initial annual revenue of $\sim$ 80,000\$/\,MW and the operation and maintenance costs (\gls{OM}) are estimated to increase with a growth rate of 6\%. The \gls{OM} costs are assumed to amount to an initial value of 2\% of the initial investment cost \gls{InvCost}. The growth rate is subject to various economic factors within the global economy, which is why a reliable prediction into the future is impossible. The rate has therefore been chosen based on historical data and recommendations by the Inter-American Development Bank~\cite{camposTimeGoesRecent2015}. Naturally, the result is very sensitive to the growth and discount rate, the reference discount rate is aligned with the growth rate to \gls{interest} = 6\%, which leads to a \gls{NPV} for a 2\,MW system of 1,016,500\,\$/\,MW over a 15-year investment horizon, as shown in \cref{tab:npv}. Changing \gls{interest} to 8\%, reduces the \gls{NPV} to 995,670\,\$/\,MW, while a much bigger increase of \gls{interest} = 20\% results in a \gls{NPV} of 885,360\,\$/\,MW. This leads to the conclusion that the investment is profitable, even for very high discount rates of up to 20\%. All economic parameters for the performed analysis are listed in tabular form in the Appendix in \cref{tab:para_econ}.

\begin{figure}[t]
    \begin{subfigure}[b]{\columnwidth}
        \centering
        \includegraphics{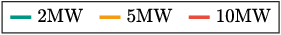}
    \end{subfigure}
    
    \vspace{0.1cm}
    
    \begin{subfigure}[b]{\columnwidth}
        \raggedleft
        \includegraphics[width=0.946\columnwidth]{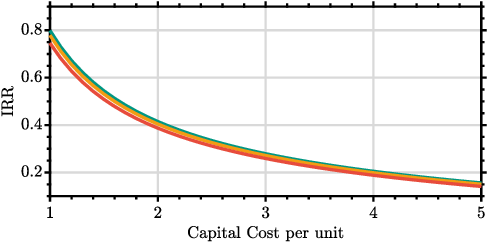}
        \caption{}
        \label{fig:IRR-capital-cost}
    \end{subfigure}
    
    \begin{subfigure}[b]{\columnwidth}
        \raggedleft
        \includegraphics[width=0.96\columnwidth]{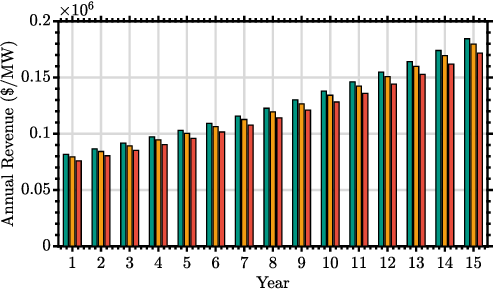}
        \caption{}
        \label{fig:revenue_yearly}
    \end{subfigure}

    \begin{subfigure}[b]{\columnwidth}
        \centering
        \includegraphics{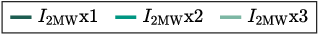}
    \end{subfigure}

        \vspace{0.1cm}
            
    \begin{subfigure}[b]{\columnwidth}
        \raggedleft
        \includegraphics[width=0.94\columnwidth]{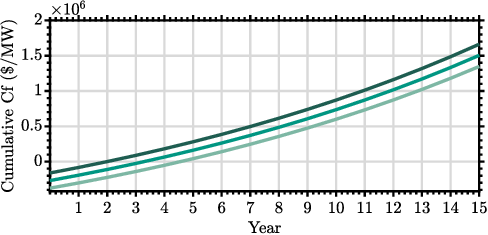}
        \caption{}
        \label{fig:cashflow_projection}
    \end{subfigure}
    
    \caption{The figures display the \gls{IRR} for a variation of the capital cost of a \gls{B2B} + \gls{LFT} by a factor of 1 to 5 in a), the projected yearly revenue for a 15 year horizon based on the annual revenue in b). The top legend applies to figure a) and b) while figure c) has its own legend. c) shows the cumulative cash flow forecast for a 2\,MW system and variations of the investment cost \gls{InvCost}.}
    \label{fig:economics_PJM}
\end{figure}

The second metric to be considered in the long-term analysis is the \gls{IRR}. The \gls{IRR} over a range of investment costs, as illustrated in \cref{fig:IRR-capital-cost}, emphasizes the economic viability of the \gls{ASG} system, as in the presented scenario of \gls{DG} aggregation in the PJM region. Scaling the investment cost for the \gls{ASG} system in the range of 1 to 5 times the base investment cost always yields a minimum IRR value of more than 0.13. This shows that the results are reliable even for higher costs of production or material. Thus, even with investment costs five times higher, an interest rate of 13\% is required to make the investment more profitable than the \gls{ASG}, according to the assumptions of this study.
The \gls{IRR} is determined as the rate that brings the \gls{NPV} equal to zero, instead of the previously predefined discount factor \gls{interest}. It is calculated numerically by solving the following equation:
\begin{equation}
    0 = \sum_{t=0}^{T} \frac{\gls{Cashflow}(t)}{(1 + \gls{IRR})^t}
\end{equation}

Both previously introduced metrics are based on future cash flows, the sum of the total revenue of the simulated year (initially $\sim$80,000\,\$/\,MW). To forecast the potential long-term revenue for the next 15 years, the rather conservative revenue growth rate of 6\% is applied~\cite{camposTimeGoesRecent2015}. The yearly cash flow (Cf) is calculated by subtracting investment cost (\gls{InvCost}) and operation and maintenance cost (\gls{OM}) from the annual revenue ($\Sigma$\gls{Creg}(t)), as in:
\begin{equation}
   \gls{Cashflow}(t) = \Sigma\gls{Creg}(t) - \gls{InvCost}(t) -  \gls{OM}(t)\label{eq:cashflow}
\end{equation}
\gls{OM} is estimated as a percentage of the capital investment (2\,\%), increasing with the capital growth rate, and \gls{InvCost} is only included as a negative cash flow in the first year of the project.
The resulting revenue per MW increases to up to 184,000\,\$ displayed in \cref{fig:revenue_yearly}, and is used to calculate the resulting cash flow during the 15-year period. 
The cumulative cash flow in \cref{fig:cashflow_projection} displays the forecast for a 15-year investment period. It includes variations in the initial investment cost (\gls{InvCost}), shown in dark green, green, and light green lines. These lines represent factors of 1, 2, and 3 times the base investment cost of 137,500\$/\,MVA or 137.5\$/\,kVA. This assumes that the investor has to buy a full \gls{B2B} system in addition to a traditional \gls{LFT} if it is not already available. These capital costs are on the conservative side, Huber and Kolar estimated the material cost for a full SST to 52\$/\,kVA back in 2014, where the cost for power electronics decreases with their maturity~\cite{huberVolumeWeightCost2014}. Nonetheless, the exact cost figures are a result of confidential discussions within industrial and academic working groups, which is why a diverse analysis for a range of investment costs of up to 3 times the base 135\$/\,kVA is offered. 
Analysis of the cumulative cash flow in \cref{fig:cashflow_projection} evidently shows that the system breaks even after at least 5 years for all the shown cases, but could break even after only 2 years.

\begin{table}[t]
    \centering
    \caption{IRR, NPV and Revenue for 1x Capital Cost.}
    \label{tab:npv}
    \begin{tabular}{l c c c}
    \toprule
        $P_{\gls{ASG}}$ & IRR & NPV (per MW)& Revenue 09/23-08/24 (per MW)\\ \midrule
        \color{kit-green}$2\,MW$\color{black} & 0.80  & 1,016,500\,\$ & 81,578\,\$  \\
        \color{kit-yellow}$5\,MW$\color{black}  & 0.78  & 986,220\,\$ & 79,439\,\$ \\
        \color{kit-red}$10\,MW$\color{black}  & 0.75  & 936,090\,\$ & 75,897\,\$  \\
        \bottomrule
    \end{tabular}
\end{table}

%% file: sections/05_discussion.tex
\section{Discussion and Perspective}\label{sec:disscusion}

Globally, experts agree on the need for more renewable energy facilities, leading to power systems with reduced mechanical inertia and increased frequency volatility. This shift requires enhanced transmission infrastructure to transport clean energy from remote locations to demand centers and updates to aging systems. The evolving technological landscape presents an opportunity to adopt \gls{SST}-like power electronic systems in substations as alternatives or supplements to traditional transformers.

This study demonstrates the technical merits of a fast-frequency response facilitated by an aggregator based on an \gls{SST}-like power electronic system~\cite{waldApplicationsServicesSolidState2025}. 
Still, there remain open questions and investigations to be answered and performed, including the integration of one and/or multiple \gls{ASG} systems in large and more complex power systems with high penetration levels of inverter-based resources.
The main section of the study, the economic analysis, finds the \gls{ASG} to be a promising, market-competitive asset and filled a long-lasting gap within the research landscape regarding the economic viability of \gls{SST}-like systems, also known as energy routers or smart substations. The technical motivation for the introduction of an increased number of power electronic systems to manage the power system is extensive; one major bottleneck for system operators or private investors has always been the question of economic viability. Despite the fact that the analysis suggests the financial viability of the \gls{ASG} technology, based on a combination of American and German standards and concepts, it is expected that the changing nature of the grid will naturally create a better financial environment and most importantly compatible regulatory frameworks. This is due to the likely increase of ancillary service prices within the foreseeable future due to the expected continued trend of reduced inertia and a possible reduction of the capital cost as power electronics technologies continue to mature. However, the increase in the prices of ancillary services will not be indefinite, as the \gls{FFR} capacity may grow accordingly and lead to more competitive pricing. Nonetheless, recent developments show that the transformation of the power system could take a very long time (perhaps decades) to reach the worldwide emission reduction targets. Thus, it is concluded that the time scales in which a \gls{ASG} system providing \gls{FFR} breaks even are much shorter than the general transformation time scales of the power system.
Furthermore, providing a long-term prediction for the future trajectory and timeline of \gls{FFR} markets is beyond the scope of this work.

It is evident that the economic viability of an \gls{ASG} system necessitates a certain set of market tariffs, policies, and regulatory frameworks. The lack of such could act as constraints, barring novel concepts, such as the \gls{ASG} from implementation and commercialized use. Illustrated by the results presented in this paper, the operation of the \gls{ASG} system in the PJM market is only financially profitable with the use of a German standard for \gls{DG} units that allows the aggregation of downstream \gls{DG} and the exploitation of their power flexibility without any additional communication infrastructure. This concept would not be feasible within PJM nor in Germany. On the European side, the \gls{ENTSO-E} regulatory framework for frequency regulation does not offer a way for smaller and/or fast-acting units to participate in any fast-frequency response service, as PJM offers RegD. The finest granularity for a frequency response service is the \gls{FCR}, bidding for 4-hour windows~\cite{fernandez-munozFastFrequencyControl2020}. 

In the case of PJM, the granularity of ancillary markets is much finer, offering 5-minute remuneration windows with signals in a 2-second time resolution. On the American side, on the other hand, the droop-activated regulation does not exist, therefore, whether as an aggregator or a \gls{VPP}, coordination of behind-the-meter resources without additional communication is hardly feasible at the moment. The aggregation process, as required by regulations such as FERC Order No. 2222, which could be easily and physically attained by operating \glspl{DG} within clusters of asynchronous grids, as discussed earlier, now necessitates direct communication and interaction with behind-the-meter resources~\cite{federalenergyregulatorycommissionfercOrderNo22222021}. This diminishes overall performance due to additional delays and decreases the probability of consumer participation if it is an opt-in situation rather than inherently built into behind-the-meter resources. If all aspects are accounted for communication and cloud services for data and edge control management, the economic analysis will be very different, in addition to concerns around cyber-security and data privacy, which are beyond the interest of this paper, but worth noting. 

In addition to the fast-frequency response service provision explored in this study, the physical, asynchronous aggregation, and clustering of the power system sections can yield the following advantages:
\begin{itemize}
    \item Allowing intentional islanding during emergencies and limiting the distance to which a disturbance is propagated.
    \item Decoupling the grid into smaller clusters and microgrids that can be managed autonomously - with the aim, e.g., to reach net zero emission, net zero power import, or enforce different grid codes.
    \item The asynchronously connected grid can support the upstream grid - frequency/voltage support ~\cite{shahOnlineVoltvarControl2016,chenImpactSmartTransformer2021,waldVirtualSynchronousMachine2024}.
    \item It can establish demand-side management exploiting voltage and frequency sensitive nodes to offer power flexibility~\cite{taoInvestigationFrequencyDependency2023,courcelleMethodsComparisonLoad2023,taoPotentialFrequencybasedPower2022}.
\end{itemize}

It is argued that market regulations and tariffs also play a decisive role in the financial mechanism for cost recovery and benefit allocation. In the current American energy market, there are two general ways to recover asset investment capital and network upgrade expenditures. For generation assets, investors recover their costs through revenue in the wholesale market. For reliability assets, e.g., transformers and transmission lines, the investors recover their costs through a guaranteed return rate over a period of time. 
%However, the recovery of costs for a \gls{ASG} system lies within the market regulation and tariffs on whether it recognizes this asset as a generation asset, since it can bid in the ancillary market, or a reliability asset, since it is effectively competing with technologies such as a transformer. 
However, the cost recovery for a \gls{ASG} system depends on its classification, either as a generation asset, which participates in the ancillary market, or as a reliability asset competing with technologies such as transformers. A third path would be a hybrid model, similar to that of large-scale Energy Storage Systems which are considered Non-wire Alternativ assets. The answer to this question and the extent to which it is allowed to recover from each revenue path lies within the regulations and tariffs of the energy market in which it operates. 

On the flip side of revenues and profits, a question looms about the losses and penalties. In addition to the allocation of benefits and compensation to the asset owners and its trickle down to individual household participants, market regulation needs to determine how the losses and disciplinary actions for violations are distributed. Besides the financial losses, the liability and penalties for matters that have an adverse impact on market operations such as defaulting on power, availability commitment, or physical damages need to be determined.

Finally, the importance of socioeconomic aspects of costumers' direct participation in the mechanisms that underlie grid reliability and power market is highlighted. A careful consideration should be given to whether it would succeed through voluntary programs or it should become mandatory. This concern arises from the fact that each consumer is not only the beneficiary of the grid services itself, but also a burden to it. Answering this question falls into the territory of anthropology and sociology, thus, although worth pondering, it is beyond the scope of this paper. 

%% file: sections/06_conclusion.tex
\section{Conclusion}\label{sec:conclusion}
This paper presents an integration study for the recently developed asynchronous grid concept into power systems. The asynchronous grid connection in this study is deployed to physically aggregate distributed generation units, exploiting their local control response to a fundamental grid frequency deviation. This approach eliminates the need for using telecommunication channels. Computer simulations demonstrated the improved dynamic response for a power system as a result of fast power injection by those aggregated resources. The subsequent market analysis shows the economic viability of investment in this technology as the projected revenue well surpasses its capital cost of design and installation, after a maximum of five years. Finally, the critical role of market regulation and tariffs in the adoption of this technology, as the modern grids evolve, is discussed. 

Whilst this paper presents a proof of concept, future research could evaluate parallel operation of multiple asynchronous grid connections and their interaction as well as their scalability in larger systems. In addition, a benchmark comparison with competing technologies such as virtual power plants or direct demand side management is needed to further carve out the distinctive characteristics of the proposed concept. Furthermore, studying the interactions between the control systems of this technology and other digital controllers in the system, including inverter-based resources, along with their correlation to the system oscillations, particularly using  electromagnetic transient (EMT) models would be interesting.

%% file: sections/yy_appendix.tex
\section*{Appendix}\label{sec:appendix}

\begin{table}[h]
    \centering
    \caption{Parameters - AG Support Model.}
    \label{tab:para_model}
    \begin{tabular}{l l c}
    \toprule
        Parameter & Value\\ \midrule
        Frequency Propagation & $R$ & 0.02 \\ 
        Load composition & $\alpha$ & 1\\
        Proportional Gain Governor & $K_\textrm{pgov}$ & 421       \\ 
        Integral Gain Governor & $K_\textrm{igov}$ & 1287       \\ 
        VSM time constant & $T_\textrm{a}$   & 3.4   \\ 
        VSM Damping & $D_\textrm{p}$  & 6.6   \\ 
        $f \sim$ P Droop & $K_\textrm{pf}$  & 0.4   \\ 
        \bottomrule
    \end{tabular}
\end{table}

\begin{table}[h]
    \centering
    \caption{Parameters - Economic Analysis.}
    \label{tab:para_econ}
    \begin{tabular}{l l c}
    \toprule
        Parameter & Symbol & Value\\ \midrule
        Performance Score & $\rho$ & 0.5 \\ 
        Revenue Growth Rate~\cite{camposTimeGoesRecent2015} & & 6\,\%       \\ 
        O\&M Cost (Initial) & $c_{\text{OM}}$ & 2\,\% of \gls{InvCost}       \\ 
        O\&M Growth Rate & & 6\,\%   \\ 
        Discount Rate~\cite{camposTimeGoesRecent2015} & $r$  & 6\,\%   \\ 
        Investment Horizon & $T$  & 15 years   \\ 
        Base Investment Cost & $I$  & 137.5\,\$/kVA   \\ 
        \bottomrule
    \end{tabular}
\end{table}

\begin{table}[h]
    \centering
    \caption{Parameters - 9-Bus Power System Model.}
    \label{tab:ninebus_sys_para}
    \begin{tabular}{ccccccc}
    \toprule
        Bus & P (MW) & Q (MW) & Line & R (pu) & X (pu) & Bus Type\\ \midrule
        1 & & & 1-4 & & 0.058& Gen \\ 
        2 & 170 & 5 & 2-7 &  & 0.063& Gen \\ 
        3 & 80 & -10 & 3-9 &  & 0.057& Gen \\ 
        4 & & & 4-5 & 0.02 & 0.920&  \\ 
        5 & -120 & -50 & 4-6 & 0.02 & 0.920& Load \\ 
        6 & -100 & -40 & 6-9 &0.04 & 0.170& Load \\ 
        7 & & & 5-7 &0.035 & 0.161&  \\ 
        8 & -110 & -30 & 7-8 &0.08 & 0.072& Load \\ 
        9 & & & 8-9 &0.012 & 0.101&  \\ 
        \bottomrule
    \end{tabular}
\end{table}

%% file: sections/zz_bios.tex

% \end{IEEEbiography}


% % insert where needed to balance the two columns on the last page with
% %\newpage

\begin{IEEEbiography}[{\includegraphics[width=1in,height=1.25in,clip,keepaspectratio]{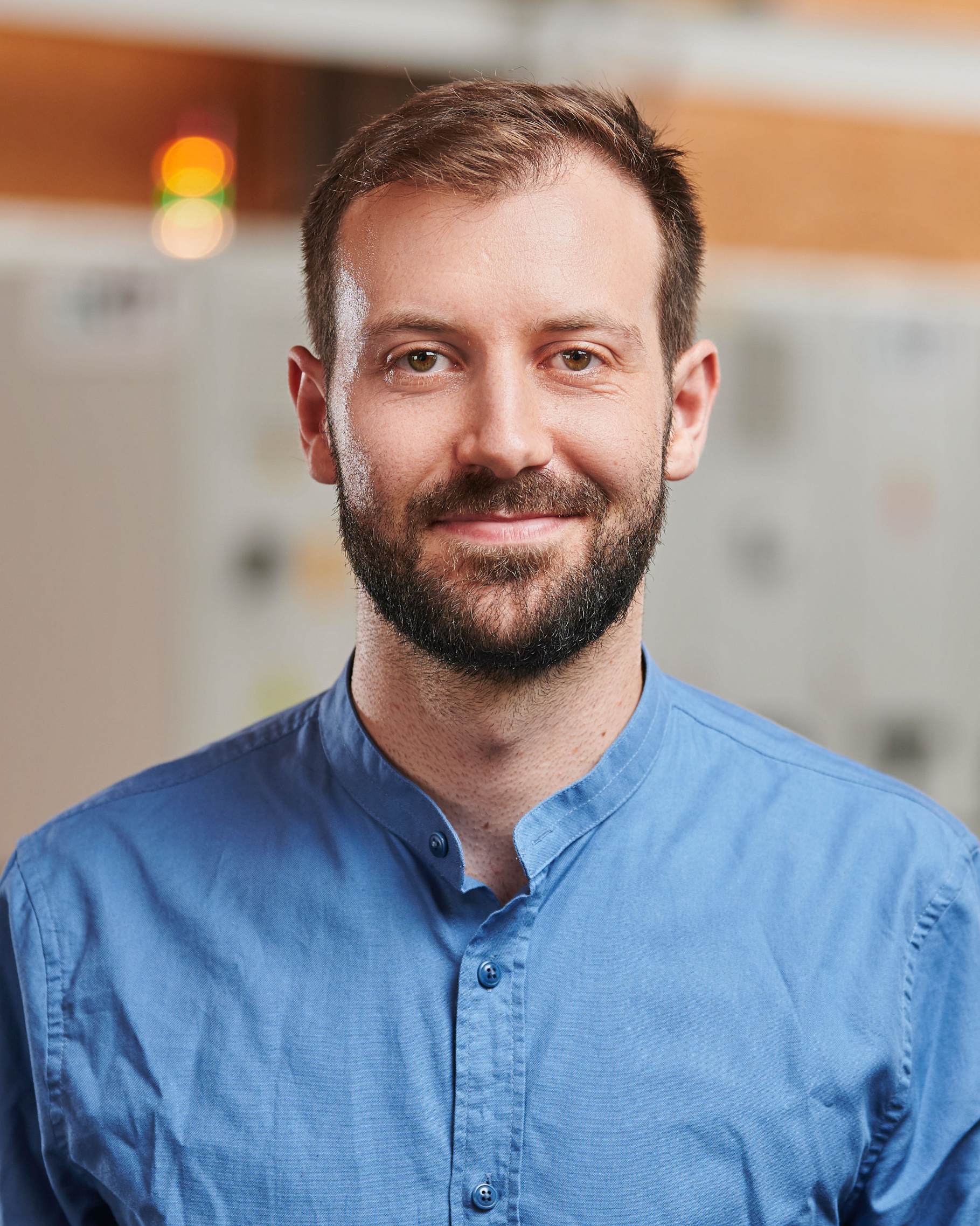}}]{Felix Wald} (GS'20) received his bachelor's degree from Berlin University of Applied Sciences in 2019 and his master's degree from the Karlsruhe Institute of Technology in 2021, in electrical engineering. Since February 2021 he is working towards his Ph.D. degree as part of the "Real Time System Integration" group and 
"Power Hardware-in-the-Loop" lab at the Institute for Technical Physics at the Karlsruhe Institute of Technology, Karlsruhe, Germany. He currently serves as secretary of the IEEE PES Taskforce on "Solid State Transformer Integration in Distribution Grids" and his research interests include Power Hardware-in-the-Loop testing and the technical and economic investigation of power electronic transformers.
\end{IEEEbiography}

\begin{IEEEbiography}[{\includegraphics[width=1in,height=1.25in]{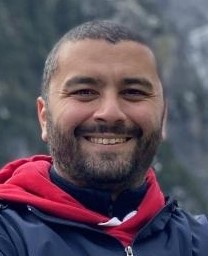}}]{Amir~Sajadi} 
(S'12, M'16, SM'19) 
is an Associate Research Professor with the Renewable and Sustainable Energy Institute, University of Colorado Boulder, Boulder, CO, USA, and a Research Affiliate with the National Laboratory of the Rockies, Golden, CO, USA. 
He received the Ph.D. degree in systems and control engineering in 2016 from Case Western Reserve University, Cleveland, OH, USA. His main research interests include modeling, planning, dynamics, control, and management of large-scale power systems to enhance the stability, security, and resiliency of energy delivery. 
\end{IEEEbiography}

\begin{IEEEbiography}[{\includegraphics[width=1in,height=1.25in,clip,keepaspectratio]{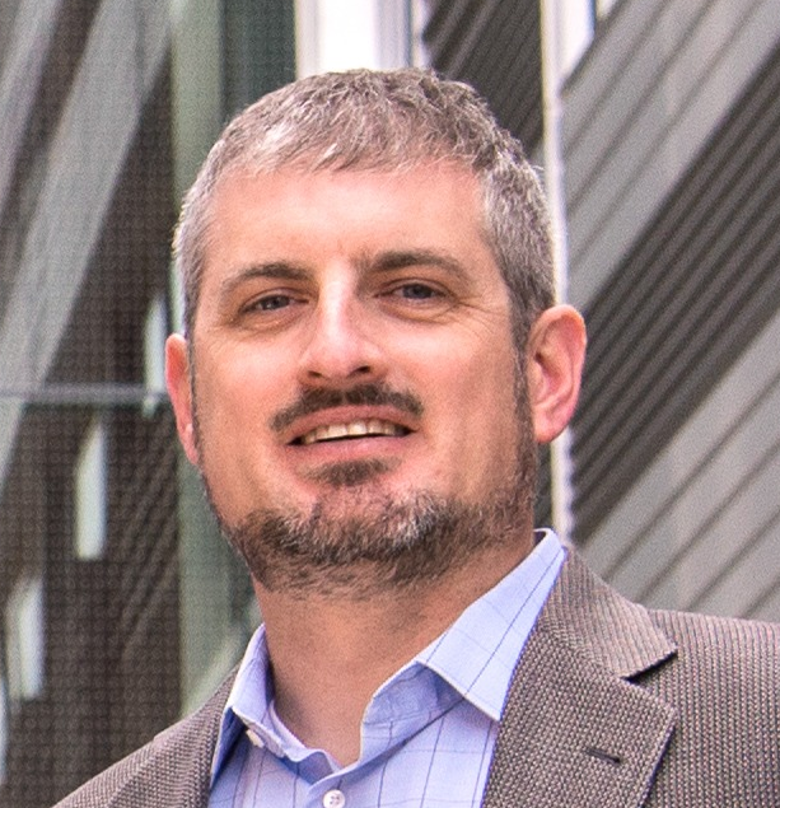}}]{Barry Mather}(SM)
received the Ph.D. degree in electrical engineering from the University of Colorado, Boulder, CO in 2010. Since 2010 he has been with the Power Systems Engineering Center at the National Laboratory of the Rockies in Golden, CO. From 2010 to 2015 he led a project focusing on the technical impacts of the integration of high-penetrations of PV in Southern California Edison’s service territory and authored the High-Penetration PV Grid Integration Handbook for Distribution Engineers. He currently leads a group of about 20 researchers focused on power electronics, system-level control, standards, and national- and state-level interconnection issues related to the integration of renewable energy sources at ever higher levels. %He is a Senior Member of the IEEE and is the chair of the IEEE Power and Energy Society’s Distribution System Analysis Subcommittee.
\end{IEEEbiography}

\begin{IEEEbiography}[{\includegraphics[width=1in,height=1.25in,clip,keepaspectratio]{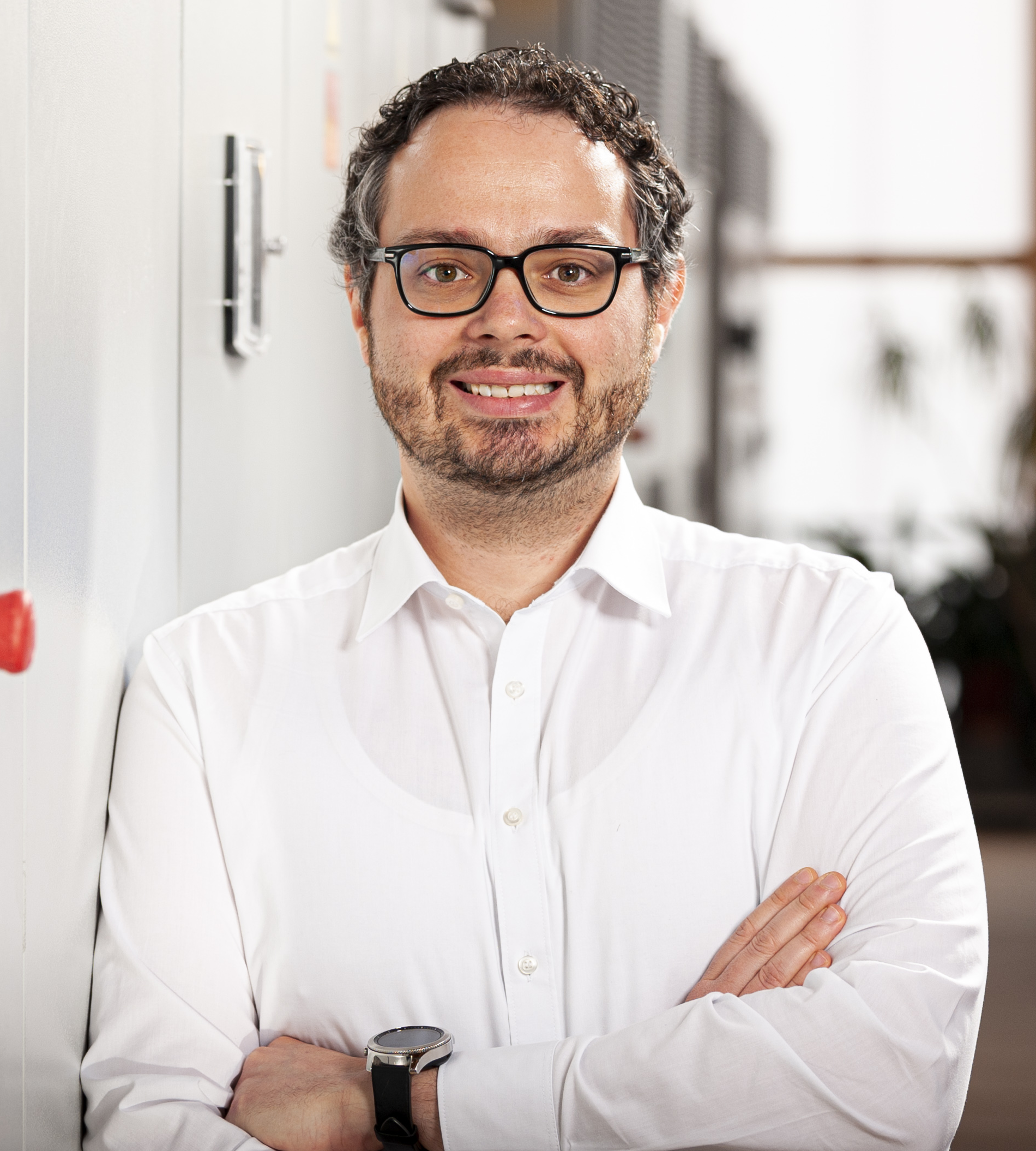}}]
{Giovanni De Carne} (S'14, M'17, SM'21) received the B.Sc. and M.Sc. degrees in electrical engineering from the Polytechnic University of Bari, Italy, in 2011 and 2013, respectively, and the Ph.D. degree from the Chair of Power Electronics, Kiel University, Germany, in 2018.
Prof. De Carne is currently full professor at the Institute for Technical Physics at the Karlsruhe Institute of Technology, Karlsruhe, Germany, where he leads the “Real Time Systems for Energy Technologies” Group and the "Power Hardware In the Loop Lab". He is currently supervising PhD students, managing academic and industrial projects, and developing multi-MW power hardware in the loop testing infrastructures for energy storage systems and hydrogen-based drives.
His research interests include power electronics integration in power systems, solid-state transformers, real-time modeling, and power hardware in the loop.
% Can be % out from here on.
He has authored/co-authored more than 100 peer-reviewed scientific papers. His research interests include power electronics integration in power systems, solid-state transformers, real-time modeling, and power hardware in the loop. In October 2023, he successfully hosted the IEEE eGrid2023 Workshop in Karlsruhe with high participation from the industry. He has been the technical program committee chair for several IEEE conferences, and associate editor of the IEEE Transaction on Power Electronics, IEEE Transactions on Power Delivery, IEEE Open Journal of Power Electronics and several other IEEE and IET journals. 
\end{IEEEbiography}


%\vfill
